\documentclass[aps,pre,twocolumn,superscriptaddress,longbibliography,nofootinbib,showkeys,showpacs]{revtex4-2}

\usepackage[utf8]{inputenc}
\usepackage{amsfonts}
\usepackage{amsmath}
\usepackage{appendix}
\usepackage{bm}
\usepackage{braket}
\usepackage{caption}
\usepackage{color}
\usepackage{comment}
\usepackage{enumerate}
\usepackage[top=2cm,left=2cm,right=2cm,bottom=2cm]{geometry}
\usepackage{graphicx}
\usepackage{hyperref}
\usepackage[utf8]{inputenc}
\usepackage{mathptmx} 
\usepackage{scalerel}
\usepackage{subcaption}
\begin{document}
\title{Unraveling Reverse Annealing: A Study of D-Wave Quantum Annealers}

\author{Vrinda Mehta\footnote{ORCID: 0000-0002-9123-7497}}
\affiliation{Institute for Advanced Simulation, J\"ulich Supercomputing Centre,\\
Forschungszentrum J\"ulich, D-52425 J\"ulich, Germany}
\author{Hans De Raedt\footnote{ORCID: 0000-0001-8461-4015}}
\affiliation{Institute for Advanced Simulation, J\"ulich Supercomputing Centre,\\
Forschungszentrum J\"ulich, D-52425 J\"ulich, Germany}
\author{Kristel Michielsen\footnote{ORCID: 0000-0003-1444-4262}}
\affiliation{Institute for Advanced Simulation, J\"ulich Supercomputing Centre,\\
Forschungszentrum J\"ulich, D-52425 J\"ulich, Germany}
\affiliation{RWTH Aachen University, 52056 Aachen, Germany}
\affiliation{AIDAS, 52425 J\"ulich, Germany}
\author{Fengping Jin\footnote{ORCID: 0000-0003-3476-524X}}
\affiliation{Institute for Advanced Simulation, J\"ulich Supercomputing Centre,\\
Forschungszentrum J\"ulich, D-52425 J\"ulich, Germany}
\begin{abstract}
    D-Wave quantum annealers offer reverse annealing as a feature allowing them to refine solutions to optimization problems. This paper investigates the influence of key parameters, such as annealing times and reversal distance, on the behavior of reverse annealing by studying models containing up to 1000 qubits. Through the analysis of theoretical models and experimental data, we explore the interplay between quantum and classical processes. Our findings provide a deeper understanding that can better equip users to fully harness the potential of the D-Wave annealers.
\end{abstract}

\date{\today}
\maketitle

\section{Introduction}

Quantum annealing has emerged as a promising approach for solving complex optimization problems by leveraging quantum mechanical effects~\cite{apolloni1988numerical,finnila1994quantum,kadowaki1998quantum,farhi2000quantum,farhi2001quantum}. Among the commercially available quantum annealers, D-Wave systems~\cite{Dwave,harris2010experimental,johnson2010scalable,johnson2011quantum,karimi2012investigating,boixo2014evidence,ronnow2014defining,albash2018adiabatic} are at the forefront, offering specialized hardware designed to solve optimization problems~\cite{dickson2012algorithmic,lucas2014ising,Venturelli_2015,bian2016mapping,ushijima2017graph,neukart2017traffic,o2018nonnegative,djidjev2018efficient,stollenwerk2019flight,willsch2020benchmarking, calaza2021garden, willsch2022tailassign, montanez2023unbalanced, montanez2024unbalanced}. Some recent studies have employed these annealers to explore complex physical phenomena in systems utilizing the annealer's architecture to simulate intricate spin interactions and study emergent behaviors~\cite{boixo2013experimental,Pudenz_2014, Lanting_2014,boixo2014evidence,boixo2014computational,bando2020probing,king2023quantum,lopez2023kagome,lopez2024quantum,king2024computationalsupremacyquantumsimulation,Vodeb25}. While considerable amount of research has focused on the forward annealing process, recent advancements have brought reverse annealing as an intriguing variant that promises to enhance the performance of quantum annealers~\cite{Dwave2017}. In reverse annealing, instead of starting in the uniform superposition state, the system is initialized in a classical state, allowing the exploration of the local energy landscape~\cite{Dwave2017}. This method has been suggested to be beneficial in guiding the system toward optimal solutions, particularly in challenging problem instances~\cite{venturelli2019reverse,passarelli2020reverse,rocutto2021quantum,Jattana2024}.

Despite the growing interest in reverse annealing, its understanding remains a subject of active debate. The extent to which quantum effects, thermalization, and classical dynamics affect the reverse annealing results is unclear. Understanding these factors is crucial for unlocking the full potential of quantum annealers. Reverse annealing has demonstrated effectiveness in certain applications~\cite{Jattana2024}, but questions about the interplay between quantum coherence, thermal noise, and dissipation still need to be answered. 
While some recent studies have addressed the amount of coherence in the dynamics of standard quantum annealing on these annealers, they primarily focused on the collective behavior of many qubits~\cite{bando2020probing,lopez2023kagome,king2023quantum,lopez2024quantum}. In contrast, this paper investigates the detailed behavior of a small number of qubits using the reverse annealing protocol.

In our previous work \cite{paper4}, we explored the effects of various control parameters offered by these annealers on the sampling probabilities of the multiple solutions of hard (for quantum annealing) 2-SAT problems. However, although these results hint at a tendency of the system to relax to equilibrium, the underlying mechanisms resulting in the observed sampling behavior are unclear. Using models like the Bloch equations and the Lindblad master equation, among others, in the present paper, we aim to delve deeper into the mechanisms that can describe such a behavior. More specifically, we experimentally study the reverse annealing feature of D-Wave for different choices of the various control parameters and incorporate features in our theoretical models that can reproduce the D-Wave results, providing deeper insights into understanding the extent to which quantum and classical processes contribute to the dynamics of reverse annealing. By shedding light on the intricate dynamics of reverse annealing, this study contributes to a deeper understanding of D-Wave quantum annealers.

The paper is organized as follows: we first provide a brief description of the methods (section~\ref{sec:methods}) employed to investigate reverse annealing and of the specific problems that we study (section~\ref{sec:probs}). We then proceed to give an overview of the key results produced by the quantum annealers (section~\ref{sec:motivation}). Next, in section~\ref{sec:simulation}, we present 
results from numerical models, designed based on the key results. This is followed by the analysis of further results from the quantum annealers (section~\ref{sec:DW}). Finally, in section~\ref{sec:conclusion} we discuss the implications of these findings, highlighting potential avenues for future research.
\section{Methods}
\label{sec:methods}

In this section, we focus on the methods and approaches used to investigate the behavior of D-Wave systems. Specifically, we limit our study to the reverse annealing protocol provided by the quantum annealers and implement the equivalent protocol in our simulations. The empirical data reported in this paper are obtained from experiments on  D-Wave Advantage\_5.4 system  at the J{\"u}lich Supercomputing Centre.

In theory, the time evolution of the D-Wave annealer is described by the Hamiltonian~\cite{Dwave}
% \begin{equation}
%     \frac{H(t)}{\hbar} = 2\pi \left( -\frac{A(s)}{2h} \sum_i\sigma_i^x + \frac{B(s)}{2h} \left[\sum_i h_i\sigma_i^z + \sum_{i>j} J_{i,j}\sigma_i^z\sigma_j^z \right] \right),
%     \label{eq:DWHamil}
% \end{equation}
\begin{align}
    \frac{H(t)}{\hbar} &= \frac{\pi A(s)}{h} H_D + \frac{\pi B(s)}{h} H_P\;, \nonumber\\
    H_D &=-\sum_i\sigma_i^x\;, \nonumber\\
    H_P &= \sum_i h_i\sigma_i^z + \sum_{i>j} J_{i,j}\sigma_i^z\sigma_j^z\;,
    \label{eq:DWHamil}
\end{align}
where $A(s)$ and $B(s)$ are expressed in GHz and are obtained by fitting simple functions to the annealing schedule data provided by D-Wave~\cite{Dwave}, and $s$ is the annealing parameter (see below).  In appendix~\ref{app:sched}, we show the D-Wave annealing schedule data and the fitted functions used in our simulations.
In reality, the time evolution under Eq.~(\ref{eq:DWHamil}) is modified by interactions with external degrees of freedom, leading to decoherence and dissipation.

The idea of the reverse annealing protocol is to start in one of the low-lying excited classical states of the problem Hamiltonian ($s=1$), and anneal backward, i.e., by decreasing the strength $B(s)$ and increasing the strength of $A(s)$ up to some reversal distance $s_r$. After an optional pause at $s_r$, the protocol then continues towards $s=1$, as in standard quantum annealing. 

The D-Wave implementation of the protocol offers several control parameters, e.g., the reverse and the forward annealing times $t_{reverse}$ and $t_{forward}$, respectively, the optional waiting time $t_{wait}$, the value of $s_r$, and the choice of the initial state. 
The reverse annealing schedule is defined by 
\begin{equation}
    s = \begin{cases}
        1-(1-s_r)\frac{t}{t_{reverse}}, &t \leq t_{reverse}\\
        s_r\;, &t_{reverse} \leq t \leq t_{reverse}\\
        s_r+(1-s_r)\frac{t-t_{reverse}-t_{wait}}{t_{forward}}\;, &t_{reverse}+t_{wait} \leq t \leq t_{end}\;,
    \end{cases}
\end{equation}
where $t_{end}=t_{reverse}+t_{wait}+t_{forward}$.

The effects of varying these parameters on the performance of reverse annealing in sampling the four degenerate ground states of the 2-SAT problems have been studied in \cite{paper4}. In this work, we focus on the behavior of the quantum annealers by changing $t_{reverse}$, $t_{wait}$, and $t_{forward}$, keeping the other parameters fixed. More specifically, we employ two different procedures described below. 

\begin{itemize}
    \item \textbf{Waiting time scans (WTS)}: In this scheme, we fix $t_{reverse}=t_{forward}=1~\mu s$, and vary $t_{wait}$. For each value of $t_{end}$, we then determine the respective probabilities $p(t_{end})$ of finding the relevant states of a given problem.
    
    \item \textbf{Annealing times scans (ATS)}: Fixing $t_{wait}=0$, in this scheme we vary $t_{reverse}=t_{forward}$. As for the other scheme, we determine the probabilities $p(t_{end})$ for the relevant energy states for each value of $t_{end}$.

\end{itemize}

In the D-Wave experiments, for each problem, we collect 4500 samples for each value of $t_{end}$ corresponding to either changing $t_{wait}$ in the WTS or changing $t_{reverse}=t_{forward}$ in the ATS.
Furthermore, in most cases shown here, the collection of the samples is carried out by sequentially submitting a given problem ten times. In doing so, we use the minor embedding feature of the annealers, which maps the given problem to a set of physical qubits. In a few other cases, we simultaneously submit many copies of the problem to the D-Wave annealer. The values of probabilities for a given state are assigned by averaging the number of times the state is sampled over the ten sequential runs in the former case,  or over the submitted number of copies in the latter.

\section{Problem description}
\label{sec:probs}
In order to gain a more general understanding of the behavior of D-Wave quantum annealers, it is beneficial to study their performance across different types of problems. To this end, we study three kinds of problems. 

\begin{itemize}
    \item \textbf{1- and 2-spin problems}: As the first class of problem Hamiltonians $H_P$'s we choose the simple instances of 1- and 2-spin problems, with fixed $h_i$'s and $J_{i,j}$. As these problems are simple, their ground state(s), first excited state(s), and the corresponding energies and degeneracies are known. Consequently, they serve as ideal problems for studying and demonstrating the behavior of the D-Wave systems. Therefore the main focus of this paper is on these problems.
    
    For the 1-spin problem, we mainly study the cases with $h_1=0$ or $h_1=0.1$. The states $\ket{\uparrow}$ and $\ket{\downarrow}$ are the two degenerate ground states of the former case. For the latter, the $\ket{\downarrow}$ state with energy $-0.1$ is the ground state, while the other state with an energy of $0.1$ is the first excited state.
    
    Moving to the 2-spin problems, we consider three problem instances.
    \begin{itemize}
        \item Instance 2S1: $h_1=h_2=-1$ and $J_{1,2}=0.95$. For this problem the $\ket{\uparrow\uparrow}$ state with energy $-1.05$ is the ground state, while the states $\ket{\uparrow\downarrow}$ and $\ket{\downarrow\uparrow}$ with energy $-0.95$ are the degenerate first excited states. State $\ket{\downarrow\downarrow}$ with energy $2.95$ is the second excited state.
        \item Instance 2S2: $h_1=h_2=-1$ and $J_{1,2}=1.00$. For this problem the states $\ket{\uparrow\uparrow}$, $\ket{\uparrow\downarrow}$, and $\ket{\downarrow\uparrow}$ are the three-fold degenerate ground states with energy $-1.00$, while the state $\ket{\downarrow\downarrow}$ with energy $3.00$ is the first excited state.
        \item Instance 2S3: $h_1=h_2=-0.95$ and $J_{1,2}=1.00$. For this problem the states $\ket{\uparrow\downarrow}$ and $\ket{\downarrow\uparrow}$ are two-fold degenerate ground states with energy $-1.00$, while $\ket{\uparrow\uparrow}$ state with energy $-0.90$ is the first excited state. State $\ket{\downarrow\downarrow}$ with energy $2.90$ is the second excited state.
    \end{itemize}
    
    \item \textbf{2-SAT problems}:
    A 2-SAT problem is defined by $M$ clauses, each consisting of two Boolean literals (a Boolean variable $x_i$ or its negation $\overline{x_i}$ for $i=1,...,N$)~\cite{paper1,paper2}. The goal is then to determine whether there exists an assignment to the variables $x_i$'s that makes each clause true, and thereby the 2-SAT problem satisfiable. As the second class of problems, we use a specially constructed set of hard (for quantum annealing) 2-SAT instances with four satisfying assignments \cite{paper4}. 
    
    To use quantum annealing for solving these problems, we map them to the Ising Hamiltonian 
    \begin{equation}
        C_{2SAT} = \sum_{\alpha=1}^M (\epsilon_{(\alpha,1)} s_{i[\alpha,1]}-1)(\epsilon_{(\alpha,2)} s_{i[\alpha,2]}-1)\;,
        \label{eq:map2SAT}
    \end{equation}
    where $i[\alpha,j]$ represents the variable $i$ that is involved in the $j$th term of $\alpha$th clause for $i=1,\dots,N$, $j=1,2$, and $\alpha=1,\dots,M$. If this variable is $x_i$ then $\epsilon_{(\alpha,j)}=1$, whereas it is its negation $\overline{x_i}$ then $\epsilon_{(\alpha,j)}=-1$.
    
    The resulting problem Hamiltonians have been found to have an exponentially increasing degeneracy of the first excited state with growing problem size \cite{paper4}.

    \item \textbf{Ferromagnetic spin chains}: As the last class of problems, we use instances of ferromagnetic spin chains, with neighbors connected by a uniform ferromagnetic coupling $J=-0.1$. The ground state of these problems is two-fold degenerate ($\ket{\uparrow\uparrow\hdots\uparrow}$ and $\ket{\downarrow\downarrow\hdots\downarrow}$) and has an energy of $J(N-1)$, where $N$ is the number of spins in the chain. 

\end{itemize}
\section{Key results}
This section showcases a few key experimental results produced by the quantum annealers which serve as the basis for the subsequent investigation.

\label{sec:motivation}
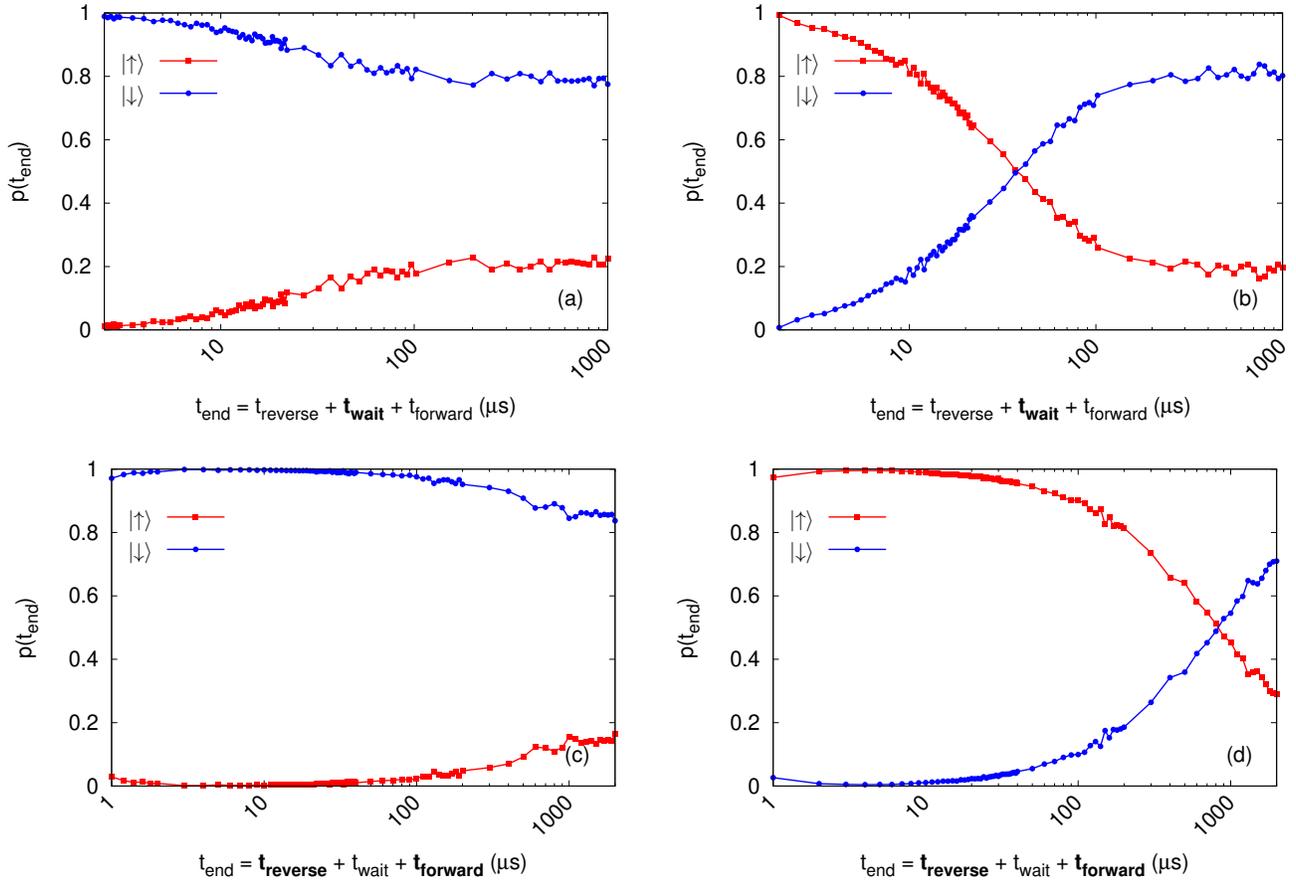
\begin{figure*}
     \begin{minipage}{0.49\textwidth}
         \centering
         \includegraphics[width=\textwidth]{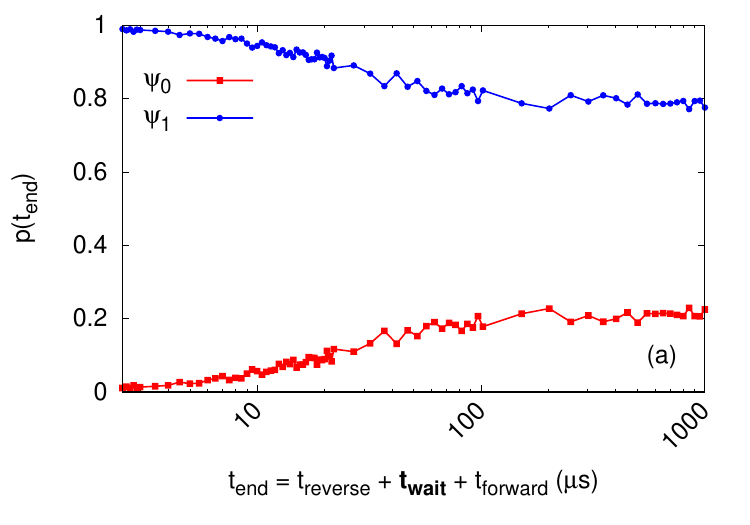}
         \put(-202, 142){\colorbox{white}{\footnotesize $\ket{\uparrow}$}}
         \put(-202, 130){\colorbox{white}{\footnotesize $\ket{\downarrow}$}}
     \end{minipage}
     \hfill
     \begin{minipage}{0.49\textwidth}
         \centering
         \includegraphics[width=\textwidth]{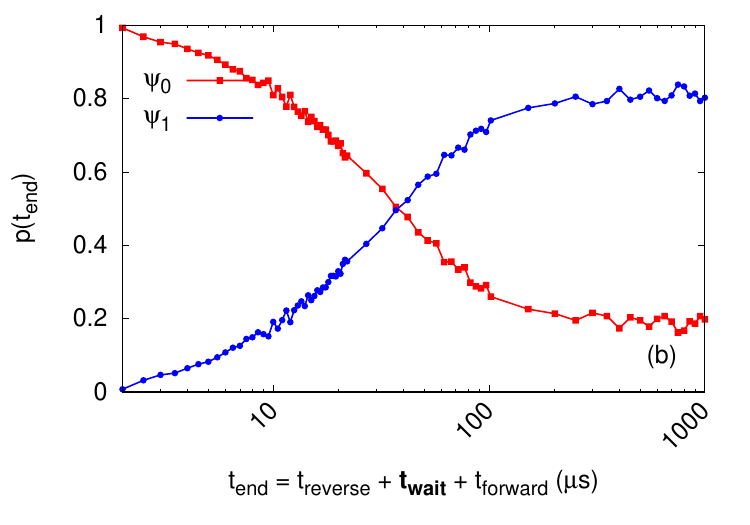}
         \put(-202, 142){\colorbox{white}{\footnotesize $\ket{\uparrow}$}}
         \put(-202, 130){\colorbox{white}{\footnotesize $\ket{\downarrow}$}}
     \end{minipage} \\
     \hfill
     \begin{minipage}{0.49\textwidth}
         \centering
         \includegraphics[width=\textwidth]{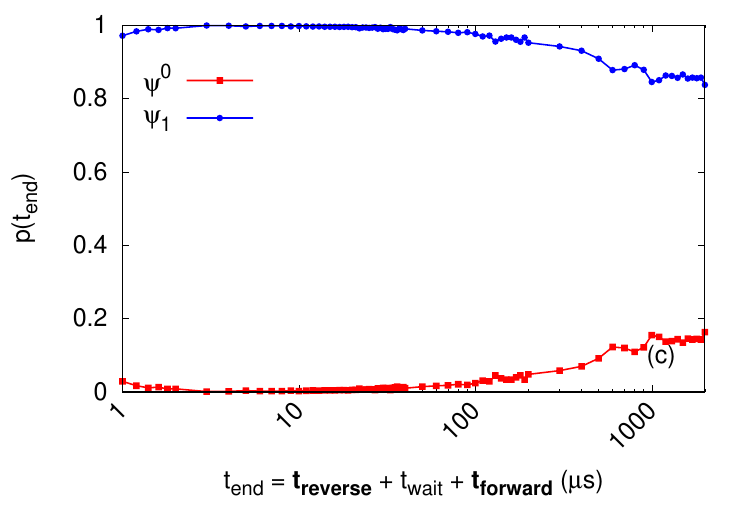}
         \put(-202, 142){\colorbox{white}{\footnotesize $\ket{\uparrow}$}}
         \put(-202, 130){\colorbox{white}{\footnotesize $\ket{\downarrow}$}}
     \end{minipage}
     \hfill
     \begin{minipage}{0.49\textwidth}
         \centering
         \includegraphics[width=\textwidth]{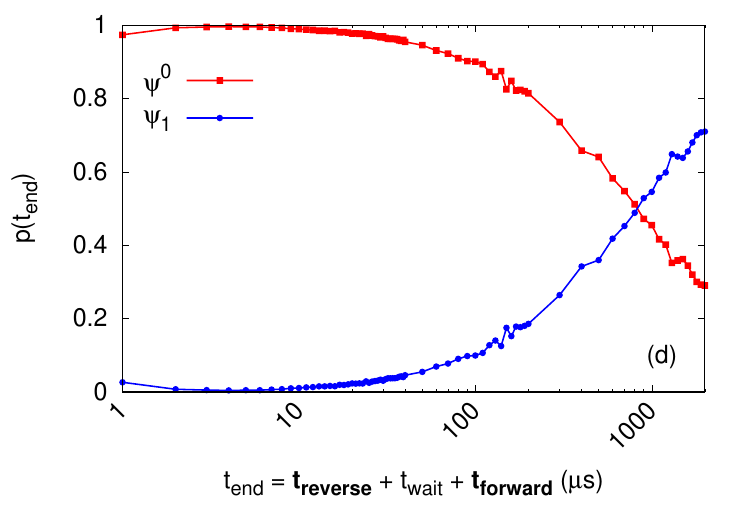}
         \put(-202, 142){\colorbox{white}{\footnotesize $\ket{\uparrow}$}}
         \put(-202, 130){\colorbox{white}{\footnotesize $\ket{\downarrow}$}}
     \end{minipage} 
     \hfill
     \caption{(Color online) (a),(b) WTS data and (c),(d) ATS data for the 1-spin problem with $h_1=0.1$ and $s_r=0.7$ obtained from the D-Wave annealer with (a),(c) $\ket{\uparrow}$ and (b),(d) $\ket{\downarrow}$ as initial states.}
    \label{fig:motiv_1spin}
\end{figure*}
We start by looking at the simplest case, a 1-spin problem, with $h_1=0.1$. The results for the WTS and the ATS are shown in Fig.~\ref{fig:motiv_1spin}. We initialize the system in the ground state $\ket{\downarrow}$ (panels (a),(c)) or first excited state $\ket{\uparrow}$ (panels (b),(d)). We note that while $p(t_{end})$ for the state that the system is initialized in starts from a value close to one for small $t_{end}$, it systematically decreases till a certain value of $t_{end}$, beyond which it seems to stabilize, especially for the WTS ($p_{\ket{\downarrow}}\approx 0.8$ for large $t_{end}$). The other state shows an opposite trend, i.e., the corresponding $p(t_{end})$ starts from a value close to zero, increases to a certain value with increasing $t_{end}$, and tends to stabilize. Furthermore, it can be seen that the rate of decrease (increase) of $p(t_{end})$ for the state which was (not) the initial state is faster for the WTS as compared to the ATS.
The results shown here are from the D-Wave Advantage\_5.4 system but other D-Wave annealers also yield a similar systematic behavior of $p(t_{end})$ (data not shown).

\begin{figure*}
     \begin{minipage}{0.33\textwidth}
         \centering
         \includegraphics[width=\textwidth]{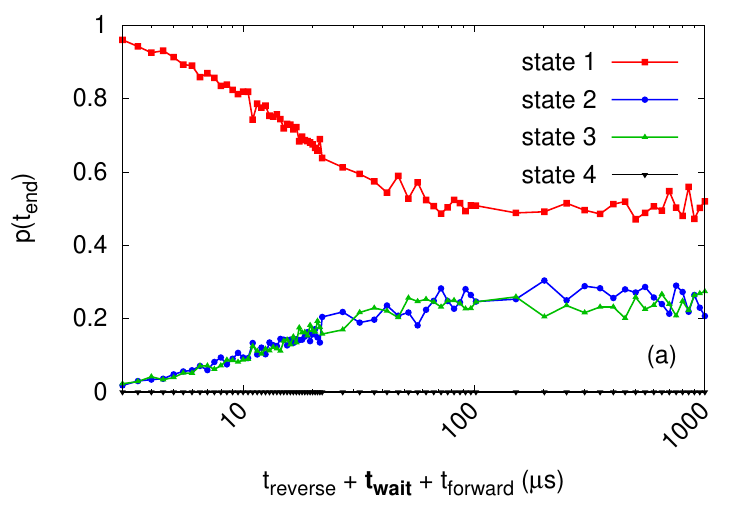}
         \put(-51, 100){\colorbox{white}{\makebox(12,2){\footnotesize $\ket{\uparrow\uparrow}$}}}
         \put(-51, 92){\colorbox{white}{\makebox(12,2){\footnotesize $\ket{\uparrow\downarrow}$}}}
         \put(-51, 84){\colorbox{white}{\makebox(12,2){\footnotesize $\ket{\downarrow\uparrow}$}}} 
         \put(-51, 75){\colorbox{white}{\makebox(12,2){\footnotesize $\ket{\downarrow\downarrow}$}}}
     \end{minipage}
     \hfill
     \begin{minipage}{0.33\textwidth}
         \centering
         \includegraphics[width=\textwidth]{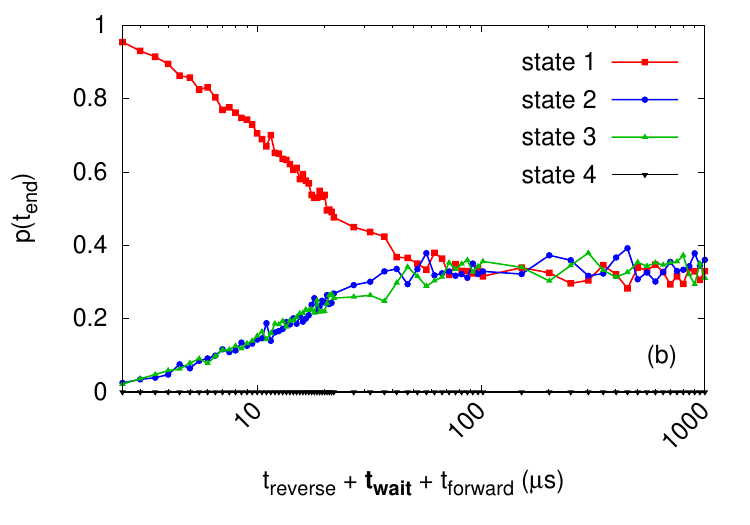}
         \put(-51, 100){\colorbox{white}{\makebox(12,2){\footnotesize $\ket{\uparrow\uparrow}$}}}
         \put(-51, 92){\colorbox{white}{\makebox(12,2){\footnotesize $\ket{\uparrow\downarrow}$}}}
         \put(-51, 84){\colorbox{white}{\makebox(12,2){\footnotesize $\ket{\downarrow\uparrow}$}}} 
         \put(-51, 75){\colorbox{white}{\makebox(12,2){\footnotesize $\ket{\downarrow\downarrow}$}}}
     \end{minipage}
        \begin{minipage}{0.33\textwidth}
         \centering
         \includegraphics[width=\textwidth]{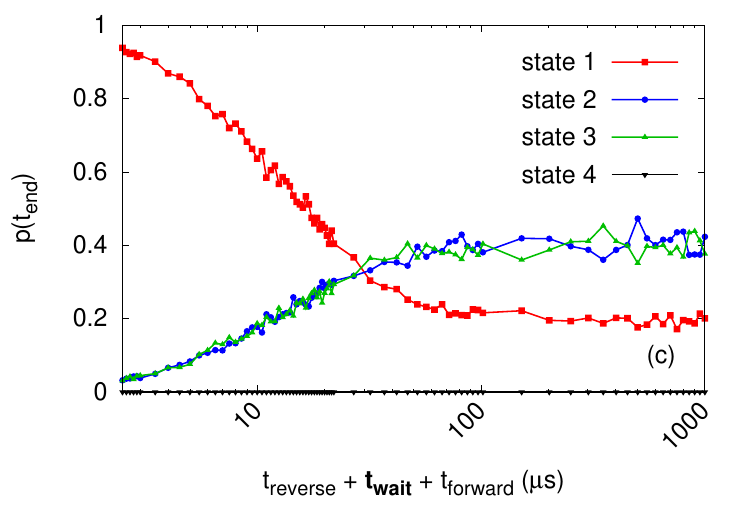}
         \put(-51, 100){\colorbox{white}{\makebox(12,2){\footnotesize $\ket{\uparrow\uparrow}$}}}
         \put(-51, 92){\colorbox{white}{\makebox(12,2){\footnotesize $\ket{\uparrow\downarrow}$}}}
         \put(-51, 84){\colorbox{white}{\makebox(12,2){\footnotesize $\ket{\downarrow\uparrow}$}}} 
         \put(-51, 75){\colorbox{white}{\makebox(12,2){\footnotesize $\ket{\downarrow\downarrow}$}}}
     \end{minipage}
     \hfill
     \caption{(Color online) WTS data for the 2-spin problem instance (a) 2S1, (b) 2S2, and (c) 2S3 obtained from the D-Wave annealer with $s_r=0.7$.}
    \label{fig:motiv_2spin}
\end{figure*}
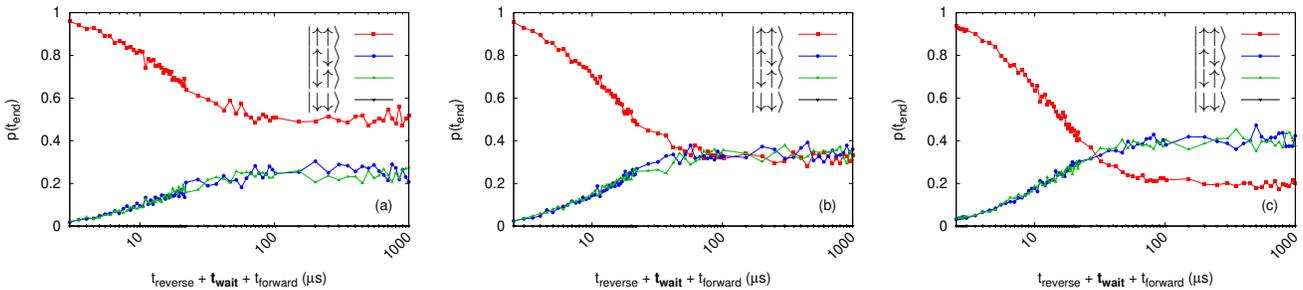
In Fig.~\ref{fig:motiv_2spin} we show the WTS for the three instances of the 2-spin problem, where we choose the state $\ket{\uparrow\uparrow}$ as the initial state. In this case, we see that $p(t_{end})$ for the $\ket{\downarrow\downarrow}$ state remains close to zero for all values of $t_{end}$. For the other states, as for the 1-spin cases, we see a similar systematic decrease (increase) in $p(t_{end})$ corresponding to the state in which the system was (not) initialized, followed by a stabilization of the $p(t_{end})$ values.

The results for the 1-spin and the 2-spin problems are, apart from minor details, reproducible and independent of the choice of the specific D-Wave system. This consistency points to a distinctive behavior of D-Wave quantum annealers, which is worth investigating further. 

These initial observations combined provide a good motivation to compare the final probabilities $p(t_{end})$ of the various states of the problem with those resulting from the equilibrium distribution.

\textbf{Conjecture:} For sufficiently large $t_{end}$ the probabilities approach their thermal equilibrium values. 

According to statistical mechanics, the equilibrium distribution is given by
\begin{equation}
    p_i^{equil} = \frac{g_i e^{-\beta E_i}}{\sum_ig_i e^{-\beta E_i}},
    \label{eq:equil}
\end{equation}
where $g_i$ and $E_i$ are the degeneracy and the energy, respectively, of the $i$th level of the problem Hamiltonian, with $E_{i+1} > E_i$, and $\beta = \eta/T$ for $\eta = h\, B(s=1)/(2 k_B)\times{10}^{9}=0.206$ and some effective temperature $T$ (expressed in kelvin). Using Eq.~(\ref{eq:equil}) and the ground state success probability obtained from the quantum annealer for the largest value of $t_{end}$ from the WTS of the 1-spin problem with $h_1=0.1$ (Fig.~\ref{fig:motiv_1spin}(a),(b)), we find $\beta=6.93$, which corresponds to an effective temperature $T=29.7$~mK, a value that is of a similar order as the cryogenic temperature of 15~mK, typical for the D-Wave annealers \cite{paper4}. Next, using the value $\beta$ obtained, we compute the ground state probabilities of the three 2-spin instances according to the equilibrium distribution. 
For instance 2S1, this results in $p_{\ket{\uparrow\uparrow}}= 0.50$, $p_{\ket{\uparrow\uparrow}}=p_{\ket{\uparrow\downarrow}}=p_{\ket{\downarrow\uparrow}}=0.25$, and $p_{\ket{\downarrow\downarrow}}=0$. For 2S2, we obtain $p_{\ket{\uparrow\uparrow}}=p_{\ket{\uparrow\downarrow}}=p_{\ket{\downarrow\uparrow}}=0.33$ and $p_{\ket{\downarrow\downarrow}}=0$, while for 2S3, Eq.~(\ref{eq:equil}) yields $p_{\ket{\uparrow\uparrow}}= 0.20$,  $p_{\ket{\uparrow\downarrow}}=p_{\ket{\downarrow\uparrow}}=0.40$, and $p_{\ket{\downarrow\downarrow}}=0$. These theoretical values closely match the probabilities obtained from the D-Wave annealer for long $t_{end}$ for these problems, thereby increasing our confidence in the conjecture that the probabilities obtained by reverse annealing on the D-Wave quantum annealers seem to relax to equilibrium probabilities for sufficiently long total annealing times.

\section{Simulation results}
\label{sec:simulation}
As discussed in the previous section, the D-Wave results are starkly different from those of ideal quantum annealing, according to which the success probability approaches one for sufficiently long annealing times. For our numerical model to capture the observed features, we need to incorporate the elements of dissipation and decoherence. To this end, we use the Gorini–Kossakowski–Sudarshan–Lindblad master equation, which approximates the Schr\"odinger dynamics of the reduced density matrix for a system interacting with an environment~\cite{gorini1976completely,breuer2002theory}. 

The Lindblad master equation reads~\cite{breuer2002theory}
\begin{align}
    \frac{d \rho(t)}{dt} &= -\frac{i}{\hbar} [H(t), \rho(t)] \nonumber\\
    &+ \frac{1}{2} \sum_j \gamma_j [2L_j\rho(t)L_j^\dagger - L_j^\dagger L_j \rho(t) - \rho(t) L_j^\dagger L_j ],
    \label{eq:lindblad}
\end{align}
where $\rho(t)$ is the density matrix of the system and $\gamma_j \geq 0 $ are the damping rates corresponding to the dissipation operators $L_j$. In general, the operators $L_j$ are linear combinations of the matrices that form a basis for the matrices operating on the Hilbert space of the system~\cite{breuer2002theory}. 

The question that remains is whether there exist choices of dissipation operators that can reproduce the results obtained from the quantum annealers for the 1-spin and the 2-spin problems presented in the previous section. In the following, we tackle these cases one by one. 

\subsection{1-spin problems: Bloch equations}
\label{sec:sim_1spin}
In general, the 1-spin Hamiltonian can be written as 
\begin{equation}
    H = -\frac{1}{2}\mathbf{B} \cdot \boldsymbol{\sigma},
    \label{eq:1spinHamil}
\end{equation}
where \textbf{B} is the applied magnetic field and $\boldsymbol{\sigma}=(\sigma^x,\sigma^y,\sigma^z)$ are the Pauli matrices. Substituting $B^x = 2\pi A(s)/h$, $B^y = 0$, and $B^z = -2\pi B(s) h_1/h$ transforms Eq.~(\ref{eq:1spinHamil}) into Eq.~(\ref{eq:DWHamil}) for a single-spin system. 

According to linear algebra, the $2 \times 2$ density matrix for the state of a spin-1/2 object can be completely expressed in the basis of the three Pauli matrices plus the unity matrix $\mathbb{I}$, i.e.,
\begin{equation}
    \rho(t) = \frac{1}{2} \left[ \mathbb{I} + \sum_\alpha S^\alpha(t) \cdot \sigma^\alpha \right],
    \label{eq:rho1spin}
\end{equation}
where $\mathbf{S(t)}=(S^x(t),S^y(t),S^z(t))$ is a vector of real numbers satisfying $\sum_\alpha(S^\alpha)^2\leq 1$ for $\alpha=x,y,z$. To solve the Lindblad equation Eq.~(\ref{eq:lindblad}) for this system, we need to find the dissipation operators $L_j$ that can produce results comparable to those obtained from the D-Wave annealers. One such choice for the dissipation operators is
\begin{align}
        L_1 &= \sigma^+ = \frac{1}{2} (\sigma^x + i \sigma^y) = \begin{pmatrix}
            0 & 1\\
            0 & 0
        \end{pmatrix},\nonumber\\
        L_2 &= \sigma^- = \frac{1}{2} (\sigma^x - i \sigma^y) = \begin{pmatrix}
            0 & 0\\
            1 & 0
        \end{pmatrix},\nonumber\\
        L_3 &= \sigma^z = \begin{pmatrix}
            1 & 0\\
            0 & -1
        \end{pmatrix}.
        \label{eq:1spindissip}
\end{align}
\begin{figure*}
     \begin{minipage}{0.49\textwidth}
         \centering
         \includegraphics[width=\textwidth]{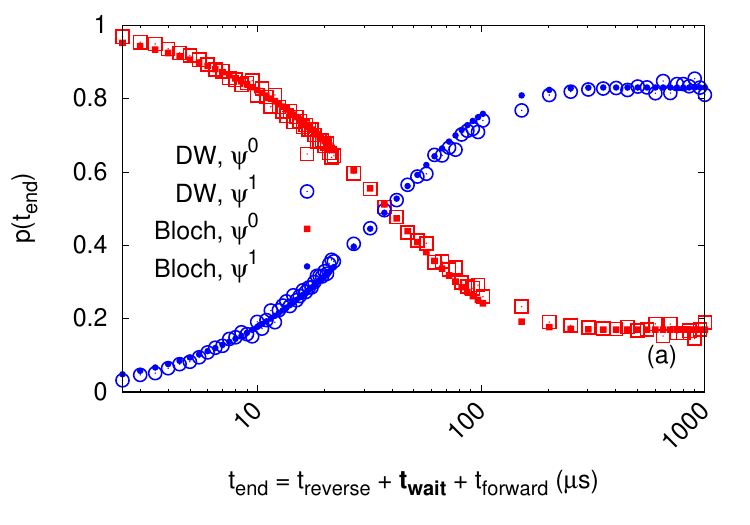}
         \put(-175, 118){\colorbox{white}{\makebox(12,6){\footnotesize $\ket{\uparrow}$}}}
         \put(-175, 106){\colorbox{white}{\makebox(12,6){\footnotesize $\ket{\downarrow}$}}}
         \put(-175, 94){\colorbox{white}{\makebox(12,6){\footnotesize $\ket{\uparrow}$}}} 
         \put(-175, 82){\colorbox{white}{\makebox(12,6){\footnotesize $\ket{\downarrow}$}}}
     \end{minipage}
     \hfill
     \begin{minipage}{0.49\textwidth}
         \centering
         \includegraphics[width=\textwidth]{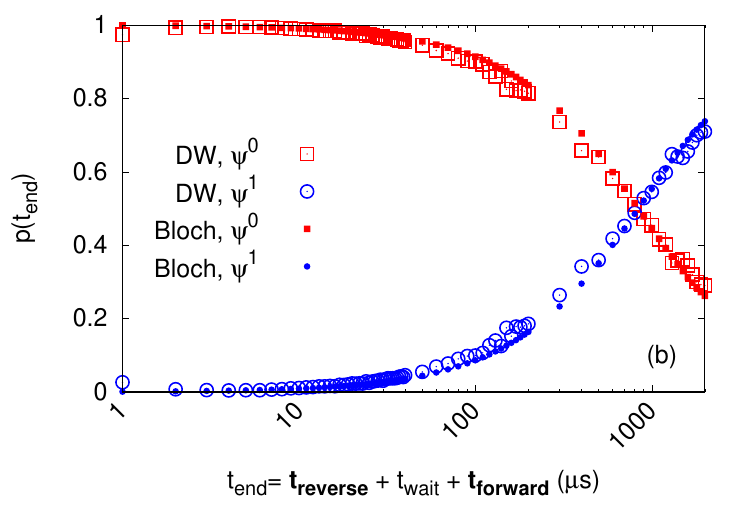}
         \put(-175, 118){\colorbox{white}{\makebox(12,6){\footnotesize $\ket{\uparrow}$}}}
         \put(-175, 106){\colorbox{white}{\makebox(12,6){\footnotesize $\ket{\downarrow}$}}}
         \put(-175, 94){\colorbox{white}{\makebox(12,6){\footnotesize $\ket{\uparrow}$}}} 
         \put(-175, 82){\colorbox{white}{\makebox(12,6){\footnotesize $\ket{\downarrow}$}}}
     \end{minipage} 
     \hfill
     \caption{(Color online) Comparison of Bloch equations simulation results with those from the D-Wave annealer for (a) WTS and (b) ATS for the 1-spin problem with $h_1=0.1$ and $s_r=0.7$. For (a) $T_1 = 41.67~\mu s$, $T_2 = 41.67~\mu s$, and $M_0 =-0.66$ while for (b) $T_1=909.09~\mu s$, $T_2=909.09~\mu s$, and $M_0=-0.66$. To emphasize that each data point from the simulations is obtained from an independent run we show each simulation data point as a marker. For improved legibility, in the subsequent figures, data points from our simulations are represented by lines through these points instead of by markers.}
    \label{fig:sim_1spin_h0.1}
\end{figure*}
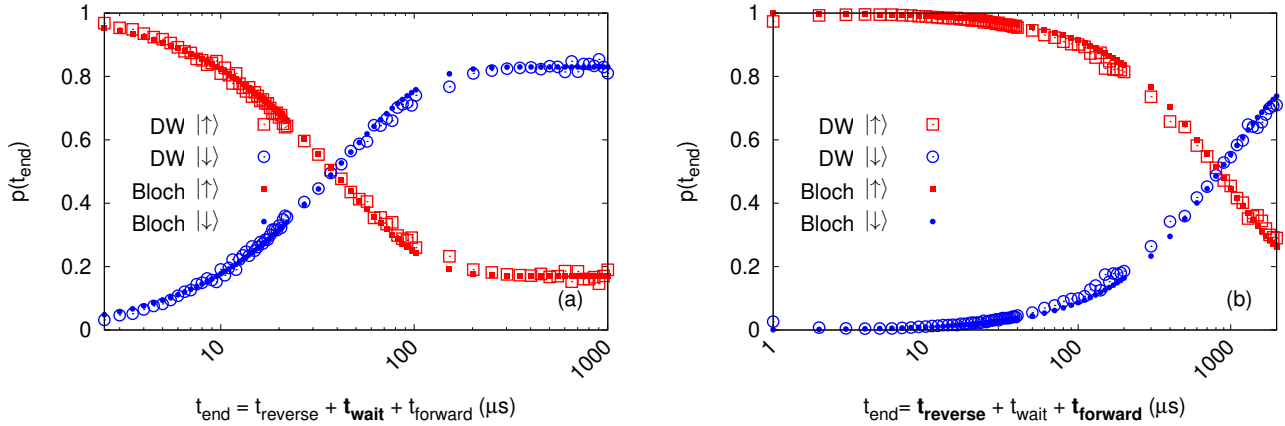
As outlined in appendix~\ref{app:1spin}, with this choice, the Lindblad equation Eq.(~\ref{eq:lindblad}) is equivalent to the Bloch equations given by
\begin{align}
           {\frac {dS^{x}(t)}{dt}}&=S^{y}(t)B^{z}(t)-S^{z}(t)B^{y}(t)-{\frac {S^{x}(t)}{T_{2}}}\nonumber\\
           {\frac {dS^{y}(t)}{dt}}&=S^{z}(t)B^{x}(t)-S^{x}(t)B^{z}(t)-{\frac {S^{y}(t)}{T_{2}}}\nonumber\\
           {\frac {dS^{z}(t)}{dt}}&=S^{x}(t)B^{y}(t)-S^{y}(t)B^{x}(t))-{\frac {S^{z}(t)-M_{0}}{T_{1}}},
           \label{eq:bloch}
\end{align}
with $T_2 = 2/(\gamma_1 + \gamma_2 + 4\gamma_3)$, $T_1=1/(\gamma_1+\gamma_2)$, $M_0=(\gamma_1-\gamma_2)/(\gamma_1+\gamma_2)$ denoting the transverse and longitudinal relaxation time and the equilibrium magnetization, respectively. Note that this choice of the dissipation operators yields $T_2 \leq 2T_1$. The numerical method used for solving Eq.~(\ref{eq:bloch}) is discussed in appendix~\ref{app:numerical}.

Figure~\ref{fig:sim_1spin_h0.1} shows a comparison of the D-Wave results with those of the WTS and ATS for the 1-spin problem with $h_1=0.1$ obtained from the simulations in which state $\ket{\uparrow}$ was chosen as the initial state. From the same figure it is evident that with the above-mentioned choice of the dissipation operators and with $T_1 = 41.67~\mu s$, $T_2 = 41.67~\mu s$, and $M_0 =-0.66$ for the WTS and $T_1=909.09~\mu s$, $T_2=909.09~\mu s$, and $M_0=-0.66$ for the ATS, the simulation can reproduce both the trend of the probability curves and the final value of the probability for large $t_{end}$ rather well. The fact that the relaxation times $T_1$ and $T_2$ obtained from fitting Bloch equations to the D-Wave data results in significantly different values for WTS and ATS only reflects that the underlying procedures are very different. Note that the relaxation times for a single qubit are obtained by procedures very different from the WTS or ATS.

\begin{figure*}
     \begin{minipage}{0.49\textwidth}
         \centering
         \includegraphics[width=\textwidth]{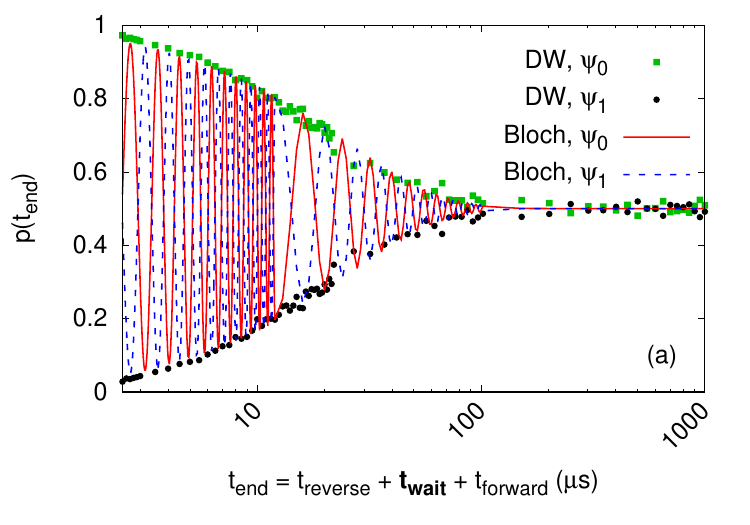}
         \put(-60, 148){\colorbox{white}{\makebox(12,6){\footnotesize $\ket{\uparrow}$}}}
         \put(-60, 136){\colorbox{white}{\makebox(12,6){\footnotesize $\ket{\downarrow}$}}}
         \put(-60, 124){\colorbox{white}{\makebox(12,6){\footnotesize $\ket{\uparrow}$}}} 
         \put(-60, 112){\colorbox{white}{\makebox(12,6){\footnotesize $\ket{\downarrow}$}}}
     \end{minipage}
     \hfill
     \begin{minipage}{0.49\textwidth}
         \centering
         \includegraphics[width=\textwidth]{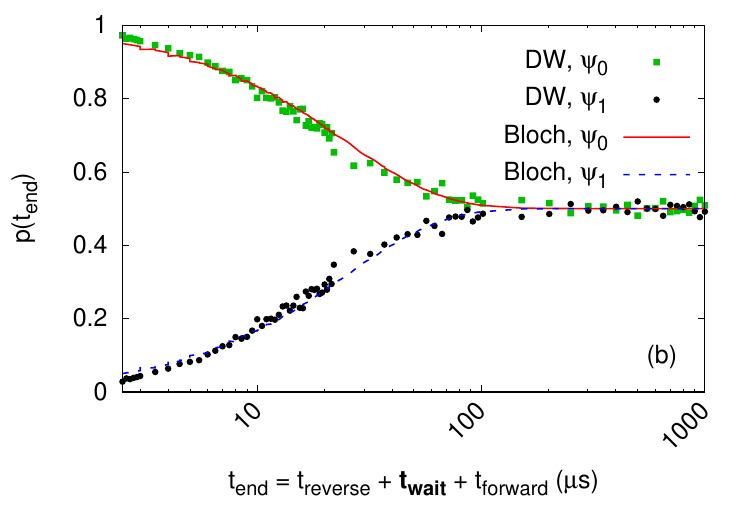}
         \put(-60, 148){\colorbox{white}{\makebox(12,6){\footnotesize $\ket{\uparrow}$}}}
         \put(-60, 136){\colorbox{white}{\makebox(12,6){\footnotesize $\ket{\downarrow}$}}}
         \put(-60, 124){\colorbox{white}{\makebox(12,6){\footnotesize $\ket{\uparrow}$}}} 
         \put(-60, 112){\colorbox{white}{\makebox(12,6){\footnotesize $\ket{\downarrow}$}}}
     \end{minipage} 
     \hfill
     \caption{(Color online) Comparison of the WTS data from Bloch equation simulations (lines) for the 1-spin problem with $s_r=0.7$ and (a) $h_1=0$ and (b) $h_1=0.001$ with the D-Wave data (markers) for the same with $h_1=0$. For both plots $T_1=25~\mu s$, $T_2=25~\mu s$, and $M_0=0$.}
    \label{fig:sim_1spin_h0}
\end{figure*}
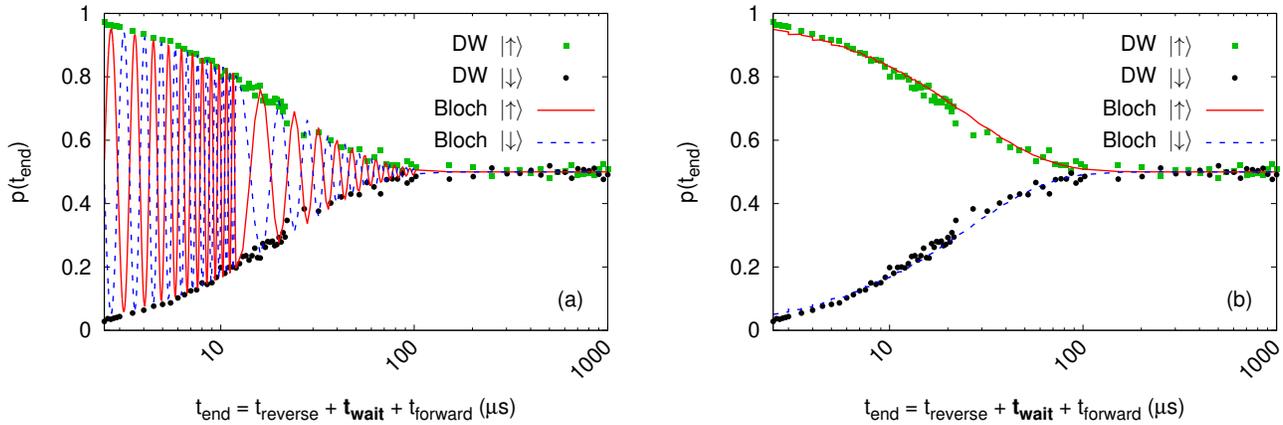

A similar treatment of the 1-spin problem with $h_1=0$ results in an oscillatory behavior between the two states of the system, as shown in Fig.~\ref{fig:sim_1spin_h0}. These oscillations are not present in the data from the quantum annealers and signal the coherent motion of the spin between the two degenerate states at $s=1$. However, slightly changing the value of the field $h_1=0.001$ eliminates these oscillations from the simulation data, and produces results that closely match those from D-Wave. As the D-Wave annealers are known to be prone to small errors in implementing the specified $h$ and $J$ values of a problem exactly, the absence of oscillations in the D-Wave data suggests that a minute amount of error in problem representation can lift the degeneracy between the energy levels.

\subsection{2-spin problems: Lindblad master equation}
\label{sec:sim_2spin}

\begin{figure*}
     \begin{minipage}{0.33\textwidth}
         \centering
         \includegraphics[width=\textwidth]{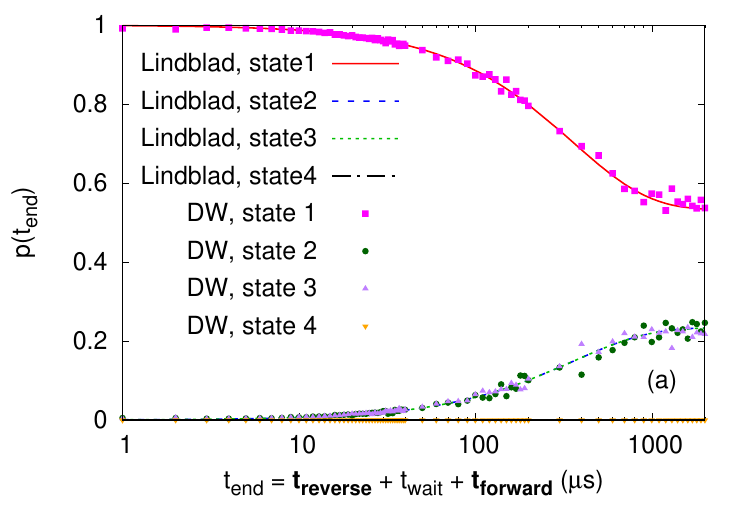}
         \put(-113, 100){\colorbox{white}{\makebox(12,2){\footnotesize $\ket{\uparrow\uparrow}$}}}
         \put(-113, 92){\colorbox{white}{\makebox(12,2){\footnotesize $\ket{\uparrow\downarrow}$}}}
         \put(-113, 84){\colorbox{white}{\makebox(12,2){\footnotesize $\ket{\downarrow\uparrow}$}}} 
         \put(-113, 75){\colorbox{white}{\makebox(12,2){\footnotesize $\ket{\downarrow\downarrow}$}}}
         \put(-113, 67){\colorbox{white}{\makebox(12,2){\footnotesize $\ket{\uparrow\uparrow}$}}}
         \put(-113, 59){\colorbox{white}{\makebox(12,2){\footnotesize $\ket{\uparrow\downarrow}$}}}
         \put(-113, 51){\colorbox{white}{\makebox(12,2){\footnotesize $\ket{\downarrow\uparrow}$}}} 
         \put(-113, 43){\colorbox{white}{\makebox(12,2){\footnotesize $\ket{\downarrow\downarrow}$}}}
     \end{minipage}
     \hfill
     \begin{minipage}{0.33\textwidth}
         \centering
         \includegraphics[width=\textwidth]{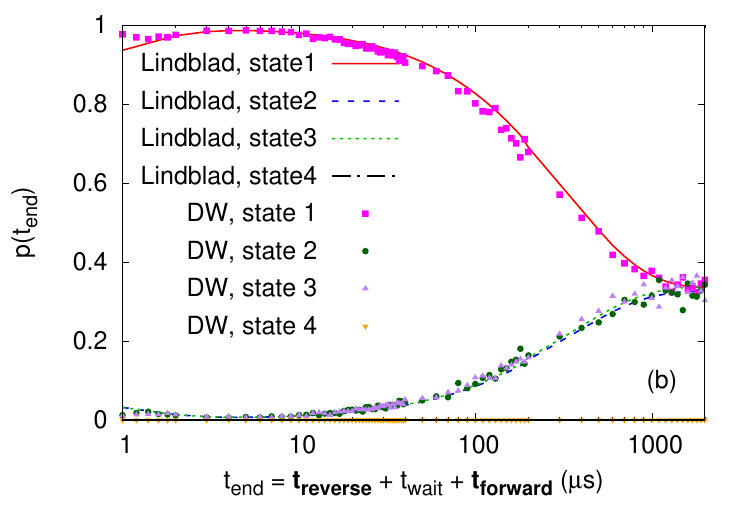}
         \put(-113, 100){\colorbox{white}{\makebox(12,2){\footnotesize $\ket{\uparrow\uparrow}$}}}
         \put(-113, 92){\colorbox{white}{\makebox(12,2){\footnotesize $\ket{\uparrow\downarrow}$}}}
         \put(-113, 84){\colorbox{white}{\makebox(12,2){\footnotesize $\ket{\downarrow\uparrow}$}}} 
         \put(-113, 75){\colorbox{white}{\makebox(12,2){\footnotesize $\ket{\downarrow\downarrow}$}}}
         \put(-113, 67){\colorbox{white}{\makebox(12,2){\footnotesize $\ket{\uparrow\uparrow}$}}}
         \put(-113, 59){\colorbox{white}{\makebox(12,2){\footnotesize $\ket{\uparrow\downarrow}$}}}
         \put(-113, 51){\colorbox{white}{\makebox(12,2){\footnotesize $\ket{\downarrow\uparrow}$}}} 
         \put(-113, 43){\colorbox{white}{\makebox(12,2){\footnotesize $\ket{\downarrow\downarrow}$}}}
     \end{minipage}
        \begin{minipage}{0.33\textwidth}
         \centering
         \includegraphics[width=\textwidth]{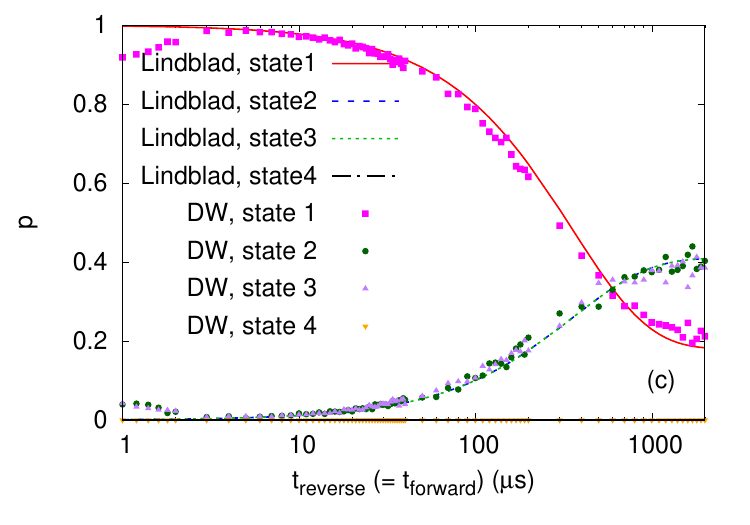}
         \put(-113, 100){\colorbox{white}{\makebox(12,2){\footnotesize $\ket{\uparrow\uparrow}$}}}
         \put(-113, 92){\colorbox{white}{\makebox(12,2){\footnotesize $\ket{\uparrow\downarrow}$}}}
         \put(-113, 84){\colorbox{white}{\makebox(12,2){\footnotesize $\ket{\downarrow\uparrow}$}}} 
         \put(-113, 75){\colorbox{white}{\makebox(12,2){\footnotesize $\ket{\downarrow\downarrow}$}}}
         \put(-113, 67){\colorbox{white}{\makebox(12,2){\footnotesize $\ket{\uparrow\uparrow}$}}}
         \put(-113, 59){\colorbox{white}{\makebox(12,2){\footnotesize $\ket{\uparrow\downarrow}$}}}
         \put(-113, 51){\colorbox{white}{\makebox(12,2){\footnotesize $\ket{\downarrow\uparrow}$}}} 
         \put(-113, 43){\colorbox{white}{\makebox(12,2){\footnotesize $\ket{\downarrow\downarrow}$}}}
     \end{minipage}
     \hfill
     \caption{(Color online) Comparison of the ATS data from D-Wave (markers) with that from Lindblad master equation simulation (lines) for 2-spin instances with $s_r=0.7$ (a) 2S1, (b) 2S2, and (c) 2S3, with dissipation rates (a)~$\gamma_1=\gamma_3=\gamma_4=\gamma_6=1.5$~Hz, $\gamma_2=0$, $\gamma_5=\gamma_7=0.6582$~Hz, (b)~$\gamma_1=\gamma_3=\gamma_4=\gamma_6=1.0$~Hz, $\gamma_2=0$, $\gamma_5=\gamma_7=0.9837$~Hz,  (c)~$\gamma_1=\gamma_3=\gamma_4=\gamma_6=0.5$~Hz, $\gamma_2=0$, $\gamma_5=\gamma_7=1.1395$~Hz.}
    \label{fig:sim_2spin}
\end{figure*}
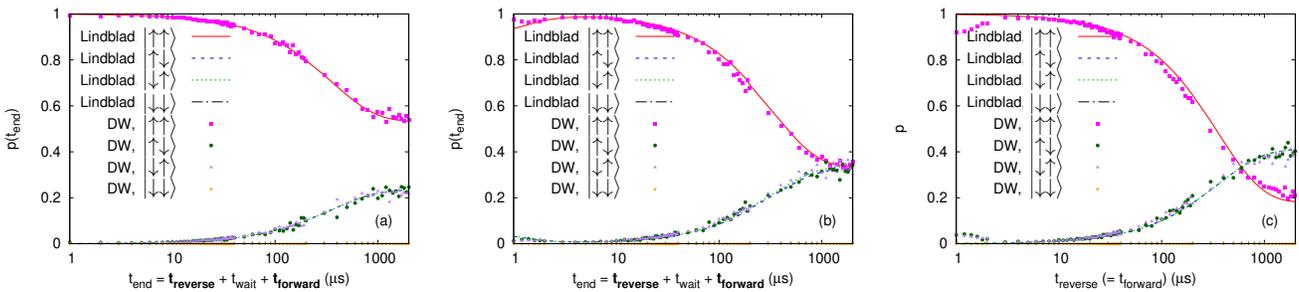

We now turn to the 2-spin problem instances discussed in the previous section. Even for such simple cases, selecting the appropriate dissipation operators demands careful deliberation. The mathematical considerations and the choice of the dissipation operators for this case are presented in appendix~\ref{app:2spin}, while the other numerical aspects for the Lindblad master equation simulation are outlined in appendix~\ref{app:numerical}.

In Fig.~\ref{fig:sim_2spin}, we show the results obtained from the simulations of the ATS for these problems, in comparison to those from the annealers. Choosing a value of $\beta$ that makes the probabilities from the equilibrium distribution Eq.~(\ref{eq:equil}) close to the probabilities $p(t_{end})$ obtained from D-Wave for the largest value of $t_{end}$, we determine ratios of the dissipation rates by setting the coherent part of the dynamics (the first term of the right-hand side of Eq.~(\ref{eq:app2spin_a})) to zero. This approach is described in more detail in the section~\ref{sec:pdotwp}. The resulting relations between the equilibrium probabilities and the dissipation rates are given by Eq.~(\ref{eq:dissiprate_equilprob}). We find that with this choice for the dissipation rates, the simulations can reproduce the behavior of the $p(t_{end})$ data produced by the D-Wave quantum annealers rather well, except for a few points corresponding to small $t_{end}$, see for instance 2S3. 

Recall that the ground state  of the 1-spin problem with $h_1=0$ is two-fold degenerate, and that if the initial state was chosen to be $\ket{\uparrow}$, the simulation data shows oscillations between the two states. For the 2-spin instance 2S2, the ground state is three-fold degenerate, and initializing the system in state $\ket{\uparrow\uparrow}$ also results in an oscillatory behavior of the corresponding probabilities (data not shown). However, as noted for the above-mentioned 1-spin problem, slightly lifting this degeneracy by setting $h_1=1.001$ and $h_2=0.999$, eliminates the oscillations. Panel (b) of Fig.~\ref{fig:sim_2spin} shows simulation results for this slightly modified version of instance 2S3.

\subsection{Larger problems: Markovian master equation}
\label{sec:pdotwp}

Our numerical results show that in cases of good agreement between the simulations and the D-Wave data, the absolute values of the non-diagonal elements of the corresponding density matrix are rather small (data not shown), suggesting that the contributions of the coherent part of the evolution can be ignored. Indeed, in the regime of interest, i.e., for $s \geq 0.7$, the effects of the transverse field produced by the $\sigma^x$ terms are negligible since according to the annealing schedule given by D-Wave $A(s)/h \leq 0.002 \ll B(s)/h$ for $s \geq 0.7$. These observations suggest that it might be worthwhile to investigate the case in which we set $A(s)=0$ and retain only the diagonal elements of the density matrix $\rho(t)$. The resulting $\rho(t)$ then commutes with Hamiltonian, and the Lindblad equations for the one- and two-spin problem reduce to
\begin{align}
    \frac{d}{dt} \begin{pmatrix}
           p_{\ket{\uparrow}}\\
           p_{\ket{\downarrow}}
         \end{pmatrix} = \begin{pmatrix}
        -\gamma_2 & \gamma_1 \\
        \gamma_2 & -\gamma_1 \\ 
    \end{pmatrix}  \begin{pmatrix}
           p_{\ket{\uparrow}}\\
           p_{\ket{\downarrow}}
         \end{pmatrix}\;,\; 
         p_{\ket{\uparrow}}+p_{\ket{\downarrow}} = 1\;, 
         \label{eq:deriv1spin1}
\end{align}
and 
\begin{align}
    &\frac{d}{dt} \begin{pmatrix}
           p_{\ket{\uparrow\uparrow}}\\
           p_{\ket{\uparrow\downarrow}}\\
           p_{\ket{\downarrow\uparrow}}\\
           p_{\ket{\downarrow\downarrow}}
         \end{pmatrix} = \begin{pmatrix}
        -\gamma_2-\gamma_5-\gamma_7 & \gamma_4 & \gamma_6 & \gamma_1 \\
        \gamma_5 & -\gamma_4 & 0 & 0 \\ 
        \gamma_7 & 0 & -\gamma_6 & 0 \\ 
        \gamma_2 & 0 & 0 & -\gamma_1
    \end{pmatrix}  \begin{pmatrix}
           p_{\ket{\uparrow\uparrow}}\\
           p_{\ket{\uparrow\downarrow}}\\
           p_{\ket{\downarrow\uparrow}}\\
           p_{\ket{\downarrow\downarrow}}
         \end{pmatrix}\,,\nonumber\\
         &p_{\ket{\uparrow\uparrow}}+p_{\ket{\uparrow\downarrow}}+p_{\ket{\downarrow\uparrow}}+p_{\ket{\downarrow\downarrow}}= 1\;, 
         \label{eq:deriv2spin1}
\end{align}
respectively. The stationary solutions for Eqs.~(\ref{eq:deriv1spin1}) and (\ref{eq:deriv2spin1}) are
\begin{equation}
    \gamma_2 p_{\ket{\uparrow}} = \gamma_1 p_{\ket{\downarrow}}\;,
\end{equation}
and 
\begin{equation}
    \gamma_2 p_{\ket{\uparrow\uparrow}} = \gamma_1 p_{\ket{\downarrow\downarrow}}\; , \gamma_5 p_{\ket{\uparrow\uparrow}} = \gamma_4 p_{\ket{\downarrow\uparrow}}\; , \gamma_7 p_{\ket{\uparrow\uparrow}} = \gamma_6 p_{\ket{\uparrow\downarrow}}\; , 
    \label{eq:dissiprate_equilprob}
\end{equation}
respectively. 

Equations~(\ref{eq:deriv1spin1}) and (\ref{eq:deriv2spin1}) are special cases of
\begin{equation}
    \frac{dP(t)}{dt} = W P(t),
    \label{eq:pdot}
\end{equation}
where $P(t) = (p_1(t), . . . , p_N (t))^T$ is a vector of non-negative elements that sum to one, and $W$ is a real-valued matrix. 
For instance, in the case of the 2-spin model, 
$P(t)$ and $W$ are given by the vector and matrix in Eq.~(\ref{eq:deriv2spin1}), respectively.

The formal solution of Eq.~(\ref{eq:pdot}) reads
\begin{equation}
    P(t) = e^{tW}P(0).
\end{equation}
As the columns of $W$ %(Eq.~(\ref{eq:W})) 
add to zero, it follows immediately that $\exp(tW)$ is a Markov matrix.

For the 2-SAT problems with four satisfying assignments \cite{paper4}, the dimension of the relevant subspace is four, and we can still use Eq.~(\ref{eq:deriv2spin1}) to study the relaxation processes.

\begin{figure*}
     \begin{minipage}{0.49\textwidth}
         \centering
         \includegraphics[width=\textwidth]{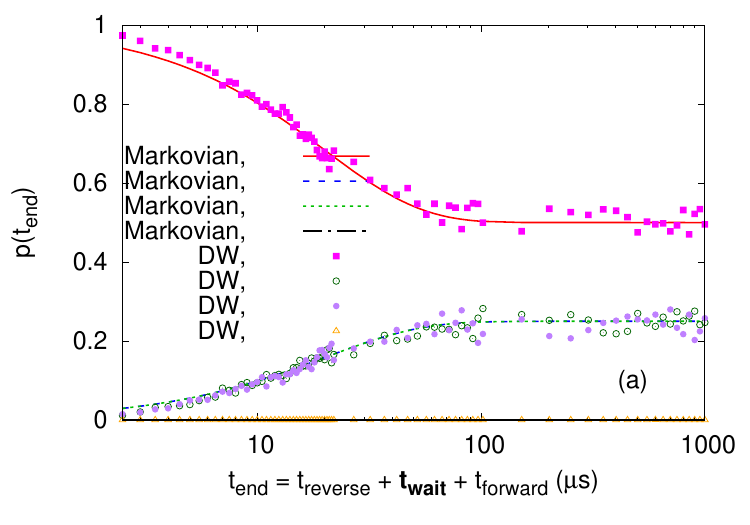}
         \put(-167, 119){\colorbox{white}{\makebox(12,2){\footnotesize $\ket{\uparrow\uparrow}$}}}
         \put(-167, 111){\colorbox{white}{\makebox(12,2){\footnotesize $\ket{\uparrow\downarrow}$}}}
         \put(-167, 103){\colorbox{white}{\makebox(12,2){\footnotesize $\ket{\downarrow\uparrow}$}}} 
         \put(-167, 95){\colorbox{white}{\makebox(12,2){\footnotesize $\ket{\downarrow\downarrow}$}}}
         \put(-167, 87){\colorbox{white}{\makebox(12,2){\footnotesize $\ket{\uparrow\uparrow}$}}}
         \put(-167, 79){\colorbox{white}{\makebox(12,2){\footnotesize $\ket{\uparrow\downarrow}$}}}
         \put(-167, 71){\colorbox{white}{\makebox(12,2){\footnotesize $\ket{\downarrow\uparrow}$}}} 
         \put(-167, 63){\colorbox{white}{\makebox(12,2){\footnotesize $\ket{\downarrow\downarrow}$}}}
     \end{minipage}
     \hfill
     \begin{minipage}{0.49\textwidth}
         \centering
         \includegraphics[width=\textwidth]{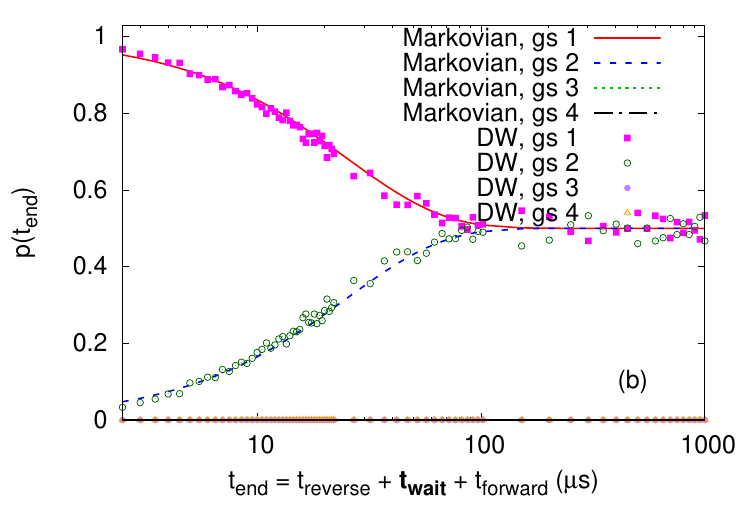}
     \end{minipage} \\
     \hfill
     \begin{minipage}{0.49\textwidth}
         \centering
         \includegraphics[width=\textwidth]{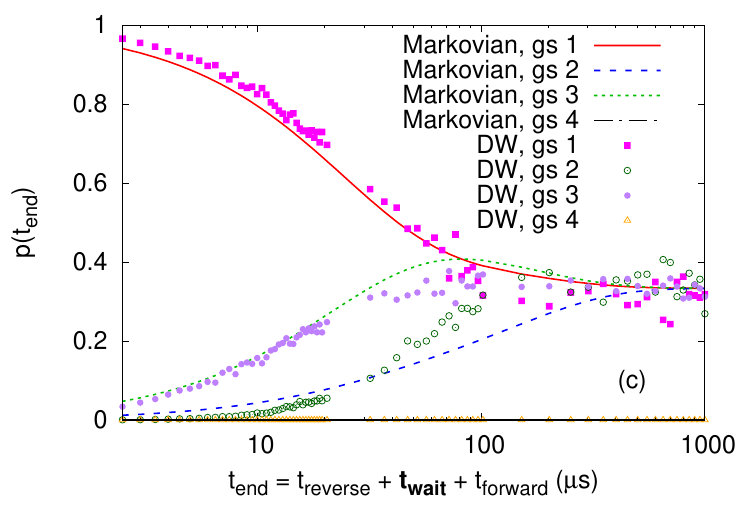}
     \end{minipage}
     \hfill
     \begin{minipage}{0.49\textwidth}
         \centering
         \includegraphics[width=\textwidth]{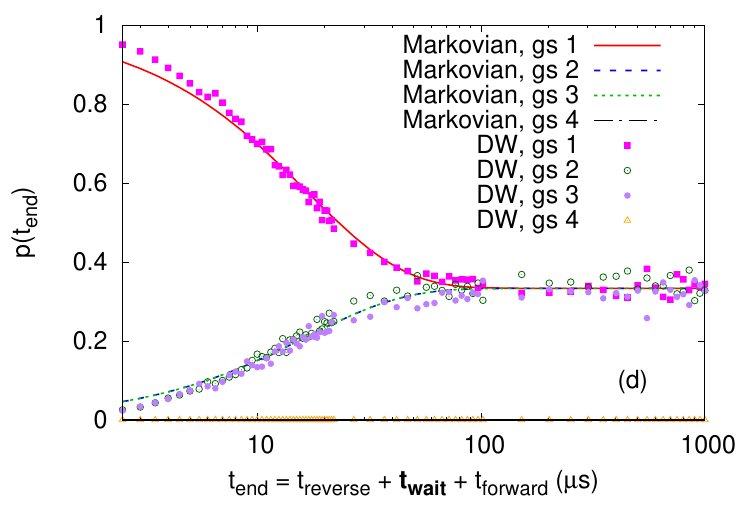}
     \end{minipage} 
     \hfill
     \caption{Comparison of the WTS data from D-Wave (markers) with that from a Markovian master equation simulation (lines) with $s_r=0.7$ for (a) instance 2S1, an instance of (b) 6-variable, (c) 12-variable, and (d) 14-variable 2-SAT problems, the last three cases with four degenerate ground states (gs 1-4) for dissipation rates (a)~$\gamma_1=\gamma_3=\gamma_4=\gamma_6=25$~Hz, $\gamma_2=0$,$\gamma_5=\gamma_7=12.5$~Hz, (b)~$\gamma_1=\gamma_2=\gamma_3=\gamma_4=\gamma_5=\gamma_6=\gamma_7=20$~Hz (c)~$\gamma_1=20$~Hz, $\gamma_2=0$, $\gamma_3=\gamma_6=\gamma_7=10$~Hz, $\gamma_4=\gamma_5=15$~Hz, (d)~$\gamma_1=\gamma_2=\gamma_3=\gamma_4=\gamma_5=\gamma_6=\gamma_7=20$~Hz.}
    \label{fig:sim_pdotwp}
\end{figure*}
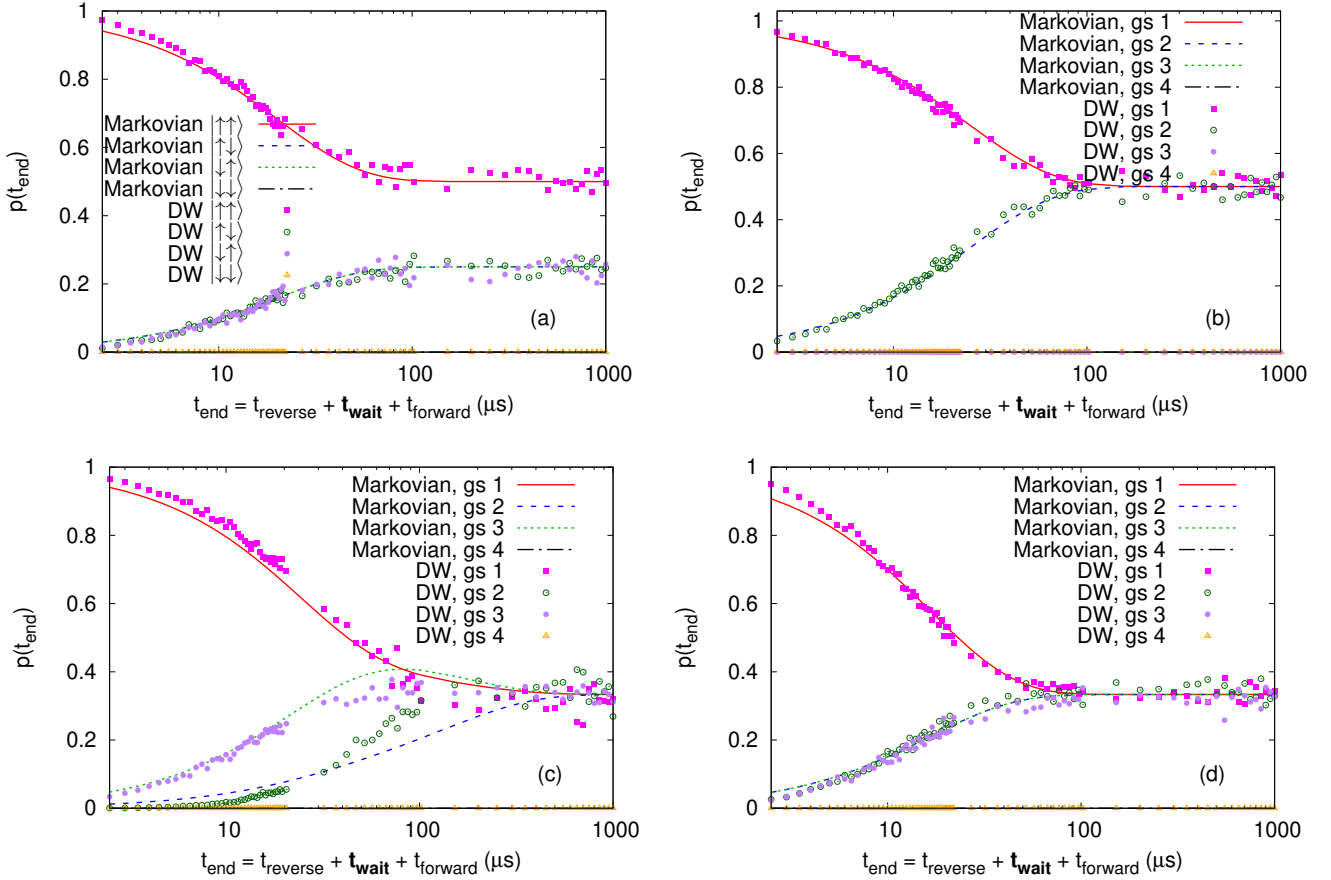

The corresponding results for the 2-spin problem, as well as for larger instances of 2-SAT problems with four satisfying assignments are shown in Fig.~\ref{fig:sim_pdotwp}. For the 2-SAT problems, this figure shows the sampling probabilities of the four known degenerate ground states of these problems (numbered in a different order than in Ref.~\cite{paper4}).

The good agreement between the numerical and the D-Wave data suggests that the salient features of the results produced by D-Wave's reverse annealing protocol can indeed be captured by this straightforward Markovian approach. 

As discussed in \cite{paper4}, for the shown instances of the 2-SAT problems, the ideal quantum annealing simulations show a suppression of the sampling probabilities of one or more ground states, a behavior that could be explained on the basis of perturbation theory. While the standard forward annealing protocol on the D-Wave annealer did not show any such suppression of any of the ground states \cite{paper4}, the suppression of these ground states using the reverse annealing protocol, as evident in Figs.~\ref{fig:sim_pdotwp}(b--d), begs for further investigation. However, it has been observed that ground states with suppressed sampling probabilities have a large Hamming distance from other ground states \cite{paper4}. This idea can be visualized by imagining the ground subspace as divided into two subspaces, one, containing states that are reachable by (a sequence of) single spin flips starting from any of the other states, and the other containing ground states that have a Hamming distance greater than one from all the states in the first subspace. Drawing an analogy to the Metropolis Monte Carlo algorithm, at very low temperatures, only degenerate ground states separated by a Hamming distance of one can be reached when starting from one of the lowest-energy configurations. Accessing other ground states requires traversing higher energy states, which becomes more likely at higher temperatures. This scenario bears resemblance to reverse quantum annealing results from D-Wave annealers, where the system remains in a regime of small quantum fluctuations, having been initialized in one of the ground states (see Figs.~\ref{fig:sim_pdotwp}(b--d)). A possible explanation for the suppression of these ground states in the D-Wave results could therefore be that quantum fluctuation plus noise sources within the system act like thermal fluctuations that are unable to couple the two ground subspaces characterized by the Hamming distance amongst the ground states.

Refocusing on the primary idea, we find that, remarkably, using the simple non-quantum Markovian description makes it possible to reproduce D-Wave results by circumventing the problem of finding the appropriate dissipation operators for the Lindblad equation for simulating these large problems, which is already a non-trivial task for the 2-spin problems. %

\section{Further D-Wave experiments}
\label{sec:DW}
From the results presented so far, we have seen the effects of varying the waiting time and the annealing times on the probability values $p(t_{end})$ for the different states of the problem. However, these WTS or ATS were performed for specific problems, and the reversal distance $s_r=0.7$.

As the next step, it is interesting to investigate how the results of these scans change when either the energy gap between the ground state and the first excited state of the problem Hamiltonian, or the reversal distance in the reverse annealing protocol, is varied. To this end, we first perform the WTS for the 1-spin problem for various values of $h_1$, fixing $s_r=0.7$. The value of the corresponding energy gap is given by
\begin{equation}
    \Delta = 2\sqrt{A^2(s)+B^2(s)h_1^2}\;,
    \label{eq:gap}
\end{equation}
at $s=s_r$. Next, setting $h_1=0.3$, we perform WTS for various values of the reversal distance $s_r$. To improve the statistics, we embed 5000 copies of the problem on the D-Wave system for each point of the scan.
\begin{figure*}
\begin{minipage}{0.49\textwidth}
         \centering
         \includegraphics[width=\textwidth]{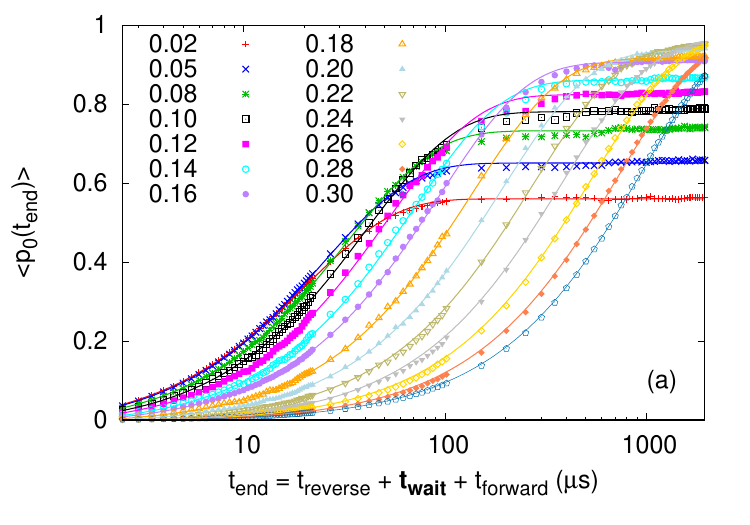}
     \end{minipage}
     \hfill
     \begin{minipage}{0.49\textwidth}
         \centering
         \includegraphics[width=\textwidth]{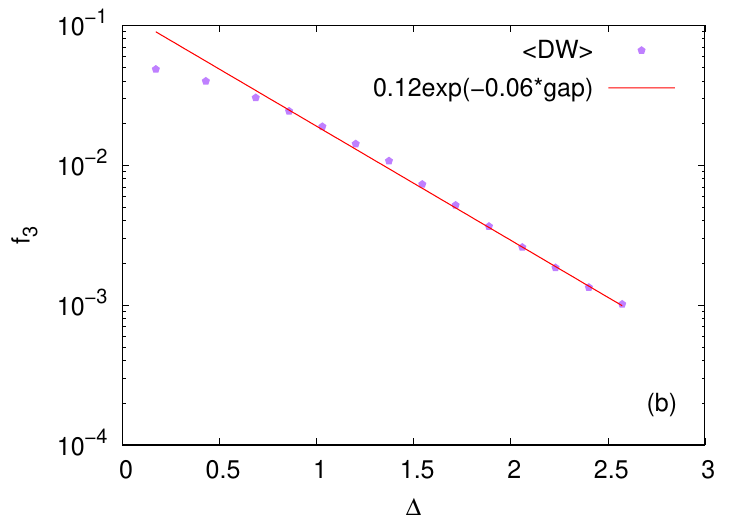}
     \end{minipage} 
     \hfill
     \\
     \begin{minipage}{0.49\textwidth}
         \centering
         \includegraphics[width=\textwidth]{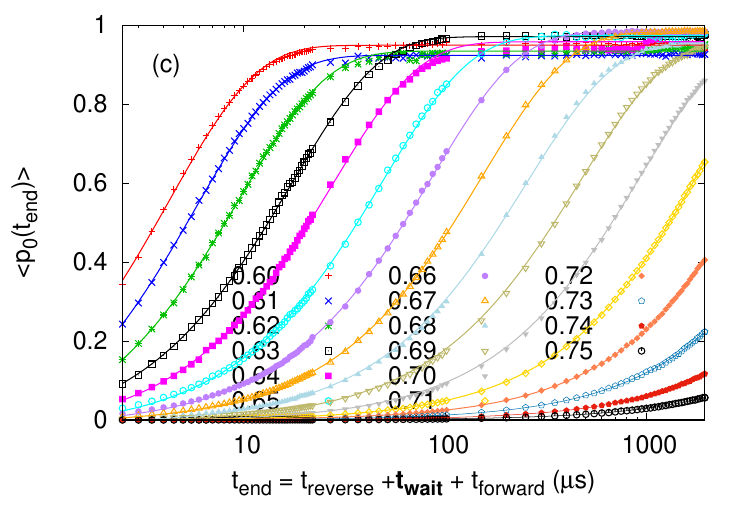}
     \end{minipage}
     \hfill
     \begin{minipage}{0.49\textwidth}
         \centering
         \includegraphics[width=\textwidth]{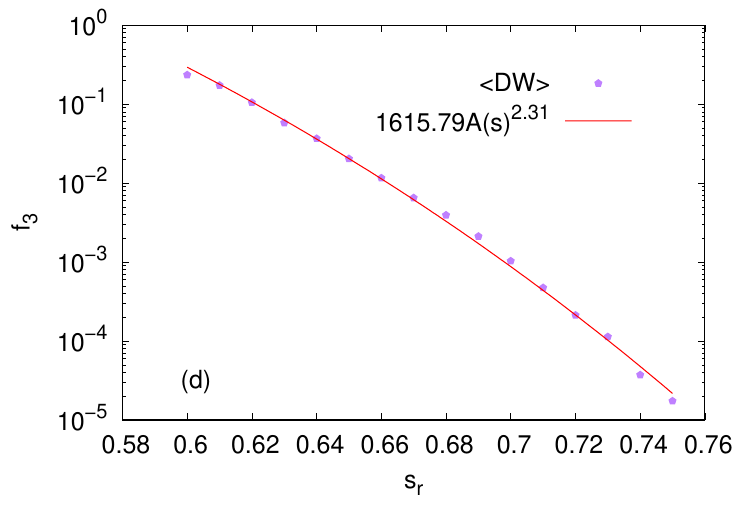}
     \end{minipage} 
     \hfill
     \caption{(Color online) (a) Data (markers) of the WTS for the mean success probability $\langle p_0(t_{end}) \rangle$ obtained by averaging 5000 copies of 1-spin problems with different $h_1$ and $s_r=0.7$ on the D-Wave annealer with the corresponding fits (lines) to $f(t_{end})=f_1(1-f_2\exp(-f_3 t_{end}))$, (b) scaling of the obtained $f_3$ as a function of energy gap $\Delta$, (c) same as (a) but for different values of reversal distance $s_r$ and $h_1=0.3$, (d) same as (b) but as a function of $s_r$.}
    \label{fig:DW_arrhenius}
\end{figure*}
We initialize the system in the first excited state, and for given values of $h_1$ and $s_r$, we perform the WTS by fixing $t_{reverse}=t_{forward}=1~\mu s$ for all the 5000 copies of the problem. \\

The results obtained from these experiments are shown in Fig.~\ref{fig:DW_arrhenius}, where panels (a) and (c) show the mean success probabilities $\langle p_0(t_{end})\rangle$ (markers) for different values of $h_1$ and $s_r$, respectively, obtained by averaging over the 5000 copies. The solid lines in these figures are obtained by fitting functions $f(t_{end}) = f_1(1-f_2\exp(-f_3t_{end}))$ to $\langle p_0(t_{end})\rangle$ obtained from the experiments for different values of $h_1$ ($s_r$) in panel (a) ((c)). Recall that each data point in these panels has been obtained from an individual run with a fixed $t_{wait}$. Therefore, it is remarkable that these points fit very well to $f(t_{end})$.

Furthermore, Fig.~\ref{fig:DW_arrhenius}(a), where $s_r$ is fixed to 0.7, shows that for $h_1 \leq 1.6$, the $\langle p_0(t_{end}) \rangle$'s saturate to values given by Eq.~(\ref{eq:equil}), in concert with the key results (section~\ref{sec:motivation}). For $h_1 > 0.16$, the maximum annealing time ($2000~\mu s$) allowed on the D-Wave systems is too short to access the regime of equilibration. In Fig.~\ref{fig:DW_arrhenius}(b) we present a plot of $f_3$ obtained from panel (a) with the energy gaps $\Delta$ calculated using Eq.~(\ref{eq:gap}), which fits well to the exponential function $0.12 \exp(-0.06 \Delta)$ for $\Delta > 1$. The exponential decrease of $f_3$ with $\Delta$ suggests that if the thus far observed systematic trend continues, increasing energy gaps between the ground state and the first excited states of the problems should postpone the attainment of the equilibrium probabilities to larger values of $t_{end}$. Such a behavior is consistent with thermal equilibration.

Next, keeping $h_1$ fixed at 0.3, we perform the WTS while varying the reversal distance $0.60 \leq s_r \leq 0.75$. Figure~\ref{fig:DW_arrhenius}(c) shows  $\langle p_0(t_{end})\rangle$ as a function of $t_{end}$ for different values of $s_r$ and suggests that the approach of the $\langle p_0(t_{end})\rangle$ to the equilibrium value given by Eq.~(\ref{eq:equil}) becomes slower with increasing values of $s_r$. In the allowed maximum annealing time limit of the D-Wave QPU of $2000~\mu s$, this saturation can only be observed up to $s_r \leq 0.68$. 

Figure~\ref{fig:DW_arrhenius}(d) shows $f_3$ as a function of $s_r$. We find that $f_3$ fits well to $1615.79 A(s_r)^{2.31}$, hinting that a stronger involvement of the transverse field leads to more fluctuations and a faster decay of $\langle p_0(t_{end})\rangle$. This observation  could also explain requiring larger dissipation rates in our simulations for the 1-spin case with $h_1=0.1$, shown in Fig.~\ref{fig:sim_1spin_h0.1}, to fit well to the D-Wave data in WTS as compared to the ATS. For the former, the system spends a larger proportion of time at a smaller $s = s_r$, resulting in a larger effective transverse field as compared to the latter where due to $t_{wait}=0$, the effective $s \geq s_r$.

\begin{figure}
    \centering
    \includegraphics[width=1\linewidth]{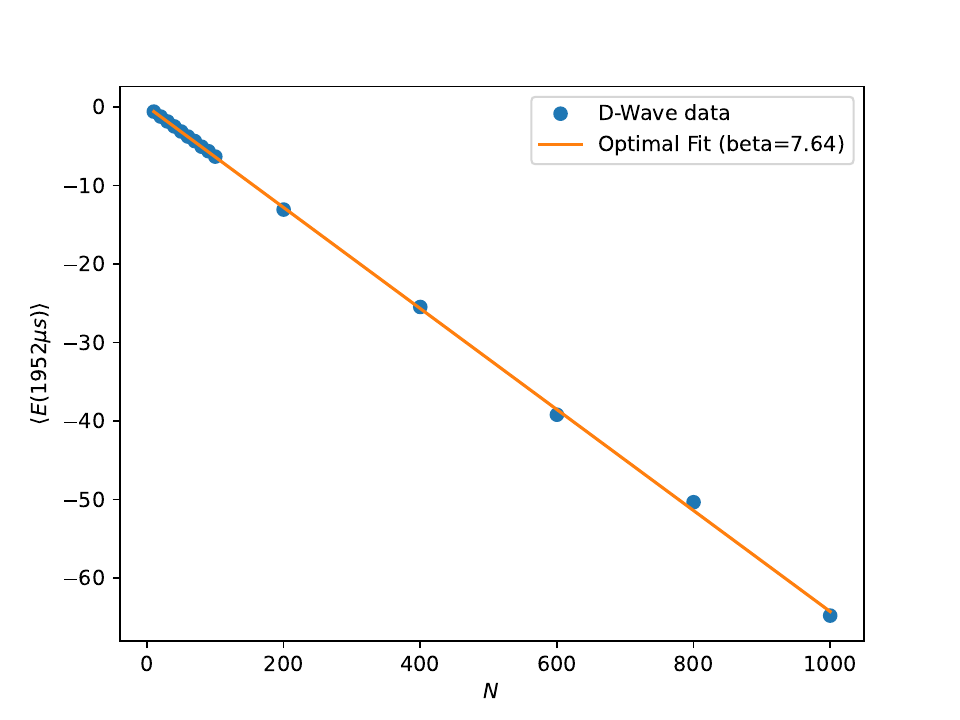}
    \caption{Mean of empirical $E(1952~\mu s)$ data obtained from the D-Wave annealer fit to Eq.~(\ref{eq:equil_energy}) with $\beta = 7.64$ corresponding to a temperature of 27~mK.}
    \label{fig:energy_optimized}
\end{figure}

Next, we study the effects of varying the problem size on the WTS using ferromagnetic spin chains for $10 \leq N \leq 1000$. For this class of problems, it is straightforward to compute the equilibrium properties. This calculation shows that the ground state probability $p_0^{equil}$ vanishes exponentially as the size of the problems increases, in concert with the corresponding D-Wave data (as shown in Fig.~\ref{fig:DW_size}(a) in Appendix~\ref{app:ferro}). To accurately estimate the ground state probability, one would therefore require an exponentially increasing number of samples as the problem size grows. Given the impracticality of this approach, a more feasible alternative is to analyze the equilibrium behavior of these problems using, for instance, the mean energy (see Fig.~\ref{fig:DW_size}(b) in Appendix~\ref{app:ferro} for the absolute value of the mean energies). In Fig.~\ref{fig:energy_optimized}, we show the mean energy $\langle E(1952~\mu s) \rangle$ as a function of the problem size $N$. We determine $\beta$ by fitting
\begin{equation}
    \langle E \rangle = \frac{\sum_i g_i E_i e^{-\beta E_i}}{\sum_ig_i e^{-\beta E_i}} = -J(N-1) \tanh \beta J,
    \label{eq:equil_energy}
\end{equation}
to the empirical data for the average energy. We find that $\beta$=7.64, which corresponds to a temperature of approximately 27~mK. This remarkably good fit strongly supports the equilibrium conjecture.

\section{Conclusion}
\label{sec:conclusion}

Using problems with known solutions, we found that for sufficiently long annealing times, the D-Wave quantum annealer samples states with frequencies approaching their thermal equilibrium probabilities. The rate at which equilibrium is attained depends on quantities like the energy gap, and the reversal distance, among other factors, limiting the possibility to approach equilibrium within the maximum annealing time admitted by the D-Wave systems. 

We have shown that with the appropriate choice of dissipation operators and rates for the Lindblad master equation, it is possible to numerically reproduce the D-Wave results for reverse annealing. Furthermore, we have shown that by ignoring in the simulations, the coherent parts, a non-quantum Markovian master equation could also reproduce the salient features of the D-Wave data. This suggests that the coherent part of the dynamics does not play a crucial role in the reverse annealing regime in D-Wave.

In light of these observations, it might be beneficial to develop better strategies for formulating/solving optimization problems using D-Wave quantum annealers.

Carrying out a similar study using standard forward annealing and the recently introduced fast annealing feature of the D-Wave quantum annealers is left for future research.

\section{Acknowledgements}
The authors gratefully acknowledge the Gauss Centre for Supercomputing e.V. (www.gauss-centre.eu) for funding this project by providing computing time through the John von Neumann Institute for Computing (NIC) on the GCS Supercomputer JUWELS~\cite{JUWELS} at the Jülich Supercomputing Centre (JSC). V.M. acknowledges support from the project JUNIQ funded by the German Federal Ministry of Education and Research (BMBF) and the Ministry of Culture and Science of the State of North Rhine-Westphalia (MKW-NRW) and from the project EPIQ funded by MKW-NRW.

\appendix
\section{Annealing schedule}
\label{app:sched}
\begin{figure}
         \centering
         \includegraphics[scale=0.7]{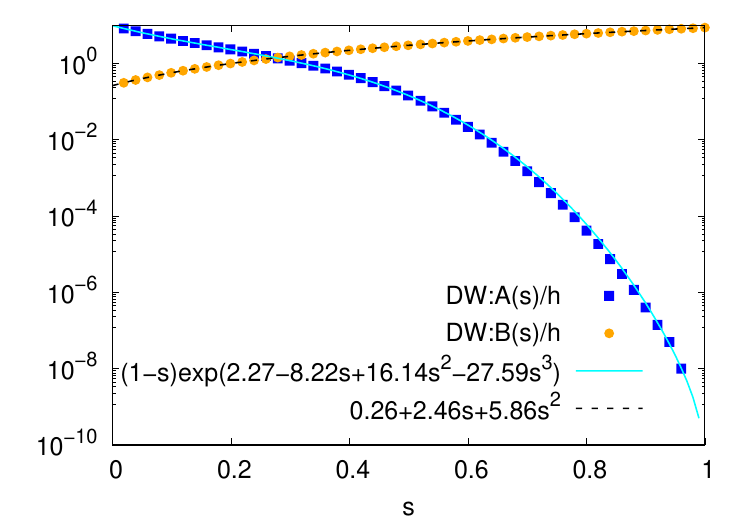}
     
     \caption{(Color online) Data for the annealing schedules  $A(s)$ and $B(s)$ provided by D-Wave, and the functions used in the simulations.}
    \label{fig:DWsched}
\end{figure}

As the D-Wave data for the annealing scheme are provided as tabulated values of $A(s)/h$ and $B(s)/h$ (in GHz), for numerical simulation, it is expedient to fit functions
\begin{align}
    A(s)/h &= (1-s) \exp(A_a + A_b s + A_c s^2 + A_d s^3),\nonumber\\
    B(s)/h &= B_a + B_b s + B_c s^2,
\end{align}
to these annealing schedule data. The values of the parameters, obtained by fitting, are shown in Fig.~\ref{fig:DWsched}.

\section{Mathematical treatment of the 1-spin case}
\label{app:1spin}

The most general Hamiltonian for a single-spin system is given by Eq.~(\ref{eq:1spinHamil}). From Eq.~(\ref{eq:rho1spin}), it follows that the eigenvalues of $\rho(t)$ are $(1+[\sum_{k=x,y,z} (S^k)^2(t)]^{1/2})/2$. Therefore, we must have $\sum_{k=x,y,z} (S^k)^2(t) \leq 1 $ for $\rho(t)$ to be a non-negative definite matrix with eigenvalues not exceeding one. From Eq.~(\ref{eq:rho1spin}) it also follows that the expectation values of the spin operators are given by 
\begin{equation}
    \langle \sigma^\alpha(t) \rangle = \mathbf{Tr}~\rho(t)\sigma^\alpha = S^\alpha(t), \alpha = x,y,z
    \;,
\end{equation}
showing that in the case at hand, knowledge of $S^\alpha(t) = \langle \sigma^\alpha(t)\rangle$ is equivalent to the knowledge of the density matrix $\rho(t)$.

We start from Eq.~(\ref{eq:lindblad}) and derive the equation of motion for the expectation values of the spin operators. This facilitates the comparison with the Bloch equations and also helps to give meaning to the damping rates that enter Lindblad equation Eq.~(\ref{eq:lindblad}). Multiplying Eq.~(\ref{eq:lindblad}) by $\sigma^l$ and computing the trace, we obtain the equations of motion of the spin expectation values (see Eq.~(\ref{eq:rho1spin})). We have
\begin{widetext}
    \begin{align}
    \frac{dS^l(t)}{dt} &= (\mathbf{S}(t)\times\mathbf{B})^l + \frac{1}{2} \sum_{j=1}^3 \mathbf{Tr}~\sigma^l [L_j,L_j^{\dagger}] + \frac{1}{4} \sum_{j,k=1}^3 \gamma_j S^k(t) [2\mathbf{Tr}~\sigma^l L_j\sigma^kL_j^{\dagger} -\mathbf{Tr}~\sigma^l L_j^{\dagger}L_j\sigma^k - \mathbf{Tr}~\sigma^l\sigma^k L_j^{\dagger}L_j] \nonumber\\
    &= (\mathbf{S}(t)\times\mathbf{B})^l + \frac{1}{2} \sum_{j=1}^3 \gamma_j \mathbf{Tr}~\sigma^l[L_j,L_j^\dagger] + \frac{1}{2} \sum_{j,k=1}^3 \gamma_j S^k(t) \mathbf{Tr}[L_j^\dagger,\sigma^l](L_j\sigma^k).
    \label{eq:app1spin_a}
\end{align}
% \end{widetext}

With the specific choice of the dissipation operators given by Eq.~(\ref{eq:1spindissip}), the right-hand side of Eq.~(\ref{eq:app1spin_a}) can be worked out analytically, yielding the Bloch equations Eq.~(\ref{eq:bloch}). It shall also be noted that the reduction of the Lindblad master equation Eq.~(\ref{eq:lindblad}) to Bloch equation Eq.~(\ref{eq:bloch}) is associated with the specific choice of the dissipation operators Eq.~(\ref{eq:1spindissip}). For a different choice of the dissipation operators, the Lindblad master equation might not reduce to the Bloch equations.

\section{Mathematical treatment of the 2-spin system}
\label{app:2spin}

For the two-spin systems, we need to choose an appropriate basis to represent $4 \times 4$ matrices. A convenient choice is the set of $4\times4$ matrices constructed by taking direct products of two matrices from the set ${\mathbb{I},\sigma^x,\sigma^y,\sigma^z}$, e.g., $e_1 = \mathbb{I}\otimes \mathbb{I}/2, e_2 = \mathbb{I}\otimes \sigma^x/2, e_3 = \mathbb{I}\otimes \sigma^y/2, e_4 = \mathbb{I}\otimes \sigma^z/2, \hdots, e_{16} = \sigma^z\otimes\sigma^z/2$. We have
\begin{equation}
    \rho(t) = \sum_{k=1}^{16} x_k(t) \mathbf{e}_k,
    \label{eq:2spinrho}
\end{equation}
where the $x_k(t)$'s are real-valued variables. Using the properties of the $\mathbf{e}_k$'s, it follows that $\mathbf{Tr}~\rho(t)=1$ implies $x_1(t) = 1/2$. The equation of motion for $x_k(t)$ for $k=2,\hdots, 16$ is found by multiplying the Lindblad equation Eq.~(\ref{eq:lindblad}) by $\mathbf{e}_k$ and computing the trace:
% \begin{widetext}
    \begin{align}
        &\frac{dx_k(t)}{dt} = \frac{d\mathbf{Tr}~\rho(t)\mathbf{e}_k}{dt} = \sum_{l=1}^{16} \left [ -i \mathbf{Tr}[H,\mathbf{e}_l]\mathbf{e}_k + \frac{1}{2} \sum_j \gamma_j \left( 2 \mathbf{Tr}L_j \mathbf{e}_l L_j^{\dagger} \mathbf{e}_k - \mathbf{Tr}L_j^\dagger \mathbf{e}_l \mathbf{e}_k - \mathbf{Tr}~\mathbf{e}_l L_j^\dagger L_j \mathbf{e}_k \right) \right] x_l(t).
        \label{eq:app2spin_a}
\end{align}
% \end{widetext}

Next, we have to choose the dissipation operators $L$'s that can describe the D-Wave data well, i.e., using which we can reproduce the trend for probabilities $p(t_{end})$'s and the stationary value that they seem to approach. The minimal choice that is found to describe this behavior is
% \begin{widetext}
\begin{align}
    &L_1 = \begin{pmatrix}
        0 & 0 & 0 & 1 \\
        0 & 0 & 0 & 0 \\
        0 & 0 & 0 & 0 \\
        0 & 0 & 0 & 0
    \end{pmatrix}\;,\; L_2 = L_1^{\mathrm{T}} 
    % \begin{pmatrix}
    %     0 & 0 & 0 & 0 \\
    %     0 & 0 & 0 & 0 \\
    %     0 & 0 & 0 & 0 \\
    %     1 & 0 & 0 & 0
    % \end{pmatrix}
    \;, 
% \end{align}
% \begin{align*}
% \nonumber \\&
    L_3 =  \begin{pmatrix}
        0 & 0 & 1 & 0 \\
        0 & 0 & 0 & 0 \\
        0 & 0 & 0 & 0 \\
        0 & 0 & 0 & 0
    \end{pmatrix}
    \;,\;
    L_4 = L_3^{\mathrm{T}}  
    % \begin{pmatrix}
    %     0 & 0 & 0 & 0 \\
    %     0 & 0 & 0 & 0 \\
    %     1 & 0 & 0 & 0 \\
    %     0 & 0 & 0 & 0
    % \end{pmatrix},
    \;,\;
% \end{align*}
% \begin{align*}
% \nonumber \\&
L_5 = \begin{pmatrix}
        0 & 1 & 0 & 0 \\
        0 & 0 & 0 & 0 \\
        0 & 0 & 0 & 0 \\
        0 & 0 & 0 & 0
    \end{pmatrix}\;,\; 
    L_6 =  L_5^{\mathrm{T}} 
    % \begin{pmatrix}
    %     0 & 0 & 0 & 0 \\
    %     1 & 0 & 0 & 0 \\
    %     0 & 0 & 0 & 0 \\
    %     0 & 0 & 0 & 0
    % \end{pmatrix}, 
    \;,\;
% \end{align*}
% \begin{align}
% \nonumber \\&
L_7 =  \begin{pmatrix}
        1 & 0 & 0 & 0 \\
        0 & -1 & 0 & 0 \\
        0 & 0 & -1 & 0 \\
        0 & 0 & 0 & 1
    \end{pmatrix}\;.
\end{align}

Using Mathematica\textsuperscript{\small\textregistered} to compute the explicit form of the equations of motion of the $x_k(t)$'s we obtain
    \begin{align}
x_1 =&\frac{1}{2}&\nonumber \\
\partial _tx_2 =&\, \frac{1}{4} \left(8 B h_1 x_3+8 B J x_{15}+\left(\gamma _1-\gamma _2-\gamma _4-\gamma _5+\gamma _6-\gamma _7\right) x_{14}-\left(\gamma _1+\gamma _2+8 \gamma _3+\gamma _4+\gamma _5+\gamma _6+\gamma _7\right) x_2\right)\nonumber \\
\partial _tx_3 =&\, \frac{1}{4} \left(-8 \left(A x_4+B h_1 x_2+B J x_{14}\right)+\left(\gamma _1-\gamma _2-\gamma _4-\gamma _5+\gamma _6-\gamma _7\right) x_{15}-\left(\gamma _1+\gamma _2+8 \gamma _3+\gamma _4+\gamma _5+\gamma _6+\gamma _7\right) x_3\right)\nonumber \\
\partial _tx_4 =&\, \frac{1}{4} \left(8 A x_3+\gamma _1-\gamma _2+\gamma _4-\gamma _5+2 \left(\gamma _1-\gamma _2-\gamma _4-\gamma _5\right) x_{16}-2 \left(\gamma _1+\gamma _2-\gamma _4+\gamma _5\right) x_7-2 \left(\gamma _1+\gamma _2+\gamma _4+\gamma _5\right) x_4\right)\nonumber \\
\partial _tx_5 =&\, \frac{1}{4} \left(8 B h_2 x_6+8 B J x_{13}+\left(\gamma _1-\gamma _2+\gamma _4-\gamma _5-\gamma _6-\gamma _7\right) x_{10}-\left(\gamma _1+\gamma _2+8 \gamma _3+\gamma _4+\gamma _5+\gamma _6+\gamma _7\right) x_5\right)\nonumber \\
\partial _tx_6 =&\, \frac{1}{4} \left(-8 \left(A x_7+B h_2 x_5+B J x_{10}\right)+\left(\gamma _1-\gamma _2+\gamma _4-\gamma _5-\gamma _6-\gamma _7\right) x_{13}-\left(\gamma _1+\gamma _2+8 \gamma _3+\gamma _4+\gamma _5+\gamma _6+\gamma _7\right) x_6\right)\nonumber \\
% \end{align*}
% \begin{align}
\partial _tx_7 =&\, \frac{1}{4} \left(8 A x_6-\gamma _2+\gamma _6-\gamma _7+\gamma _1 \left(-2 x_4-2 x_7+2 x_{16}+1\right)-2 \left(\gamma _2-\gamma _6+\gamma _7\right) x_4-2 \left(\gamma _2+\gamma _6+\gamma _7\right) \left(x_7+x_{16}\right)\right)\nonumber \\
\partial _tx_8 =&\, \frac{1}{4} \left(8 B h_1 x_9+8 B h_2 x_{11}+\left(\gamma _1+\gamma _2-\gamma _4+\gamma _5-\gamma _6+\gamma _7\right) x_{12}-\left(\gamma _1+\gamma _2+\gamma _4+\gamma _5+\gamma _6+\gamma _7\right) x_8\right)\nonumber \\
\partial _tx_9 =&\, \frac{1}{4} \left(-8 A x_{10}-8 B h_1 x_8+8 B h_2 x_{12}-\left(\gamma _1+\gamma _2-\gamma _4+\gamma _5-\gamma _6+\gamma _7\right) x_{11}-\left(\gamma _1+\gamma _2+\gamma _4+\gamma _5+\gamma _6+\gamma _7\right) x_9\right)\nonumber \\
\partial _tx_{10} =&\, \frac{1}{4} \left(8 A x_9+8 B h_2 x_{13}+8 B J x_6+\left(\gamma _1-\gamma _2+\gamma _4-\gamma _5-\gamma _6-\gamma _7\right) x_5-\left(\gamma _1+\gamma _2+8 \gamma _3+\gamma _4+\gamma _5+\gamma _6+\gamma _7\right) x_{10}\right)\nonumber \\
\partial _tx_{11} =&\, \frac{1}{4} \left(-8 A x_{14}-8 B h_2 x_8+8 B h_1 x_{12}-\left(\gamma _1+\gamma _2-\gamma _4+\gamma _5-\gamma _6+\gamma _7\right) x_9-\left(\gamma _1+\gamma _2+\gamma _4+\gamma _5+\gamma _6+\gamma _7\right) x_{11}\right)\nonumber \\
\partial _tx_{12} =&\, \frac{1}{4} \left(-8 \left(A \left(x_{13}+x_{15}\right)+B h_2 x_9+B h_1 x_{11}\right)+\left(\gamma _1+\gamma _2-\gamma _4+\gamma _5-\gamma _6+\gamma _7\right) x_8-\left(\gamma _1+\gamma _2+\gamma _4+\gamma _5+\gamma _6+\gamma _7\right) x_{12}\right)\nonumber \\
\partial _tx_{13} =&\, \frac{1}{4} \left(-8 \left(A \left(x_{16}-x_{12}\right)+B h_2 x_{10}+B J x_5\right)+\left(\gamma _1-\gamma _2+\gamma _4-\gamma _5-\gamma _6-\gamma _7\right) x_6-\left(\gamma _1+\gamma _2+8 \gamma _3+\gamma _4+\gamma _5+\gamma _6+\gamma _7\right) x_{13}\right)\nonumber \\
\partial _tx_{14} =&\, \frac{1}{4} \left(8 A x_{11}+8 B h_1 x_{15}+8 B J x_3+\left(\gamma _1-\gamma _2-\gamma _4-\gamma _5+\gamma _6-\gamma _7\right) x_2-\left(\gamma _1+\gamma _2+8 \gamma _3+\gamma _4+\gamma _5+\gamma _6+\gamma _7\right) x_{14}\right)\nonumber \\
\partial _tx_{15} =&\, \frac{1}{4} \left(-8 \left(-A x_{12}+A x_{16}+B h_1 x_{14}+B J x_2\right)+\left(\gamma _1-\gamma _2-\gamma _4-\gamma _5+\gamma _6-\gamma _7\right) x_3-\left(\gamma _1+\gamma _2+8 \gamma _3+\gamma _4+\gamma _5+\gamma _6+\gamma _7\right) x_{15}\right)\nonumber \\
\partial _tx_{16} =&\, \frac{1}{4} \left(8 A \left(x_{13}+x_{15}\right)-\gamma _5+\gamma _6-\gamma _7+\gamma _4 \left(-2 x_4+2 x_7-2 x_{16}+1\right)-2 \left(\gamma _5-\gamma _6+\gamma _7\right) x_4-2 \left(\gamma _5+\gamma _6+\gamma _7\right) \left(x_7+x_{16}\right)\right)
\;.
\end{align}

Since $\mathbf{Tr}~\mathbf{e}_1 = 2$, and $\mathbf{Tr}~\mathbf{e}_j = 0 $ for $j > 1$, from Eq.~(\ref{eq:2spinrho}) it follows that $x_1 = 1/2$ implies that the trace of the density matrix is always equal to one. 
\end{widetext}

\section{Numerical solution}
\label{app:numerical}
In matrix-vector notation, differential equations Eqs.~(\ref{eq:app1spin_a}) and (\ref{eq:app2spin_a}) take the form
\begin{equation}
    \frac{dx}{dt} = (\mathbf{C}(t) + \mathbf{D}) \mathbf{x}(t) + \mathbf{y},
    \label{eq:appNum_a}
\end{equation}
where $\mathbf{C}(t)$ is a time-dependent matrix describing the coherent motion of the spin(s), $\mathbf{D}$ accounts for the decoherence and dissipation, and the vector $\mathbf{y}$ is a time-independent source term. In the case of the single-spin $\mathbf{C}(t)$ is a $3\times 3$ skew-Hermitian matrix, $\mathbf{D}$ is a $3\times 3$ non-positive diagonal matrix, and the value of $y_3 = M_0/T_1$ determines the stationary value of the longitudinal component of the spin. For the two-spin system, excluding the trivial equation for $x_1(t)$, $\mathbf{C}(t)$ is a $15 \times 15$ skew-Hermitian matrix, and $\mathbf{D}$ is a $15\times 15$ matrix with no obvious symmetry properties, and the vector $\mathbf{y}$ with fifteen elements determines the stationary values  of the  elements of the density matrix. 

Regarding $\mathbf{C}$(t) to be piecewise constant within time intervals of duration $\tau$, i.e., assuming $\mathbf{C}(t) = \mathbf{C}_n$ for $n\tau \leq t < (n+1)\tau$, we obtain
\begin{subequations}
    \begin{align}
        \mathbf{x}((n+1)\tau) =& e^{\tau (\mathbf{C}_n+\mathbf{D})} \mathbf{x}(n\tau) + \int_0^\tau e^{(\tau-\lambda)(\mathbf{C}_n+\mathbf{D})}\mathbf{y}d\lambda
        \label{eq:appnum_ba}\\
        =& e^{\tau (\mathbf{C}_n+\mathbf{D})} \mathbf{x}(n\tau) 
        \nonumber \\
        &+ (\mathbf{C}_n+\mathbf{D})^{-1} \left( e^{\tau(\mathbf{C}_n+\mathbf{D})}-\mathbb{I} \right) \mathbf{y}.
        \label{eq:appnum_bb}
    \end{align}
    \label{eq:appnum_b}
\end{subequations}

We employ two different algorithms for solving Eq.~(\ref{eq:appnum_b}) numerically. The first method is numerical diagonalization. Since the dimension of the involved matrices here is rather small, we can compute the left-hand side of Eq.~(\ref{eq:appnum_bb}) through repeated numerical diagonalization of $\mathbf{M}=\mathbf{C}_n+\mathbf{D}$ for successive $n$, where here and in the following we suppress the subscript $n$ for notational simplicity. As $\mathbf{M}$ is a general, real-valued matrix, we have
\begin{equation}
    \mathbf{MR} = \mathbf{R} \Lambda\quad,\quad \mathbf{M} = \mathbf{R} \Lambda \mathbf{R}^{-1},
    \label{eq:appnum_c}
\end{equation}
where $\Lambda$ is a diagonal matrix with complex-valued eigenvalues $\gamma_j$ of $\mathbf{M}$ on the diagonal, $\mathbf{R}$ is the matrix of the eigenvectors of $\mathbf{M}$ as its columns, and $\mathbf{R}^{-1}$ is the inverse of $\mathbf{R}$, if it exists. If latter is true, we have
\begin{equation}
    \mathbf{x}((n+1)\tau) = \mathbf{R} e^{\tau \Lambda} \mathbf{R}^{-1} \mathbf{x}(n\tau)+\mathbf{R}\Lambda^{-1} (e^{\tau \Lambda} - \mathbb{I}) \mathbf{R}^{-1} \mathbf{y},
    \label{eq:appnum_d}
\end{equation}
where the matrices appearing in Eq.~(\ref{eq:appnum_d}) are obtained by matrix diagonalization and inversion, which is feasible for these problems, given their small size. 
Clearly, this numerically exact method can only be used if the inverse of $\mathbf{R}$ exists. 

The second alternative to solve Eq.~(\ref{eq:appnum_b}) is to make use of a product-formula algorithm based on the decomposition of matrix exponentials.  We start by approximating the contribution of the source term, represented by the last term in Eq.~(\ref{eq:appnum_ba}). Approximating the integral in Eq.~(\ref{eq:appnum_ba}) by a two-point trapezium rule we obtain
\begin{equation}
    \mathbf{x}((n+1)\tau) = e^{\tau(\mathbf{C}_n+\mathbf{D})}\mathbf{x}(n\tau) + \frac{\tau}{2} (1+e^{\tau(\mathbf{C}_n+\mathbf{D})})\mathbf{y},
    \label{eq:appnum_e}
\end{equation}
which is a second-order accurate approximation in time step $\tau$. To keep the algorithm second-order accurate, we use a second-order accurate algorithm for computing the exponential $\exp(\tau(\mathbf{C}_n+\mathbf{D}))$. 

For the single-spin system, the simplest choice for the decomposition is 
\begin{equation}
    e^{\tau(\mathbf{C}_n+\mathbf{D})} \approx e^{\tau \mathbf{A}_2/2}e^{\tau \mathbf{A}_1(n)}e^{\tau \mathbf{A}_2/2},
    \label{eq:appnum_f}
\end{equation}
where
\begin{align}
    \mathbf{A}_1(n) = \begin{pmatrix}
        0 & B^z(n) & -B^y(n)\\
        -B^z(n) & 0 & B^x(n)\\
        B^y(n) & -B^x(n) & 0
    \end{pmatrix}, \nonumber \\
    \mathbf{A}_2 = \begin{pmatrix}
        -1/T_2 & 0 & 0\\
        0 & -1/T_2 & 0\\
        0 & 0 & -1/T_1
    \end{pmatrix}.
    \label{eq:appnum_g}
\end{align}

Moving on to the two-spin system, we decompose the matrix exponential in three components, such that
\begin{equation}
    e^{\tau(\mathbf{C}_n+\mathbf{D})} \approx e^{\tau \mathbf{A}_1(n)/2}e^{\tau \mathbf{A}_2(n)/2}e^{\tau \mathbf{A}_3}e^{\tau \mathbf{A}_2(n)/2}e^{\tau \mathbf{A}_1(n)/2},
    \label{eq:appnum_h}
\end{equation}
with $\mathbf{A}_1(n)$ being the part of $\mathbf{C}_n$ with $B=0$ in Eq.~(\ref{eq:DWHamil}) and $\mathbf{A}_2(n)$ being the one with $A=0$ in Eq.~(\ref{eq:DWHamil}), and $\mathbf{A}_3=\mathbf{D}$. While the matrix exponentials of $\mathbf{A}_1 (n)$ and $\mathbf{A}_2(n)$ can be calculated analytically, we compute $\exp(\tau \mathbf{A}_3)$ by numerical diagonalization (once). The product formula approach can be applied when the inverse of $\mathbf{R}$ does not exist and, in practice, mainly provides an independent check on the numerical results.

\section{D-Wave results for ferromagnetic spin chains}
\label{app:ferro}

\begin{figure*}
\begin{minipage}{0.49\textwidth}
         \centering
         \includegraphics[width=\textwidth]{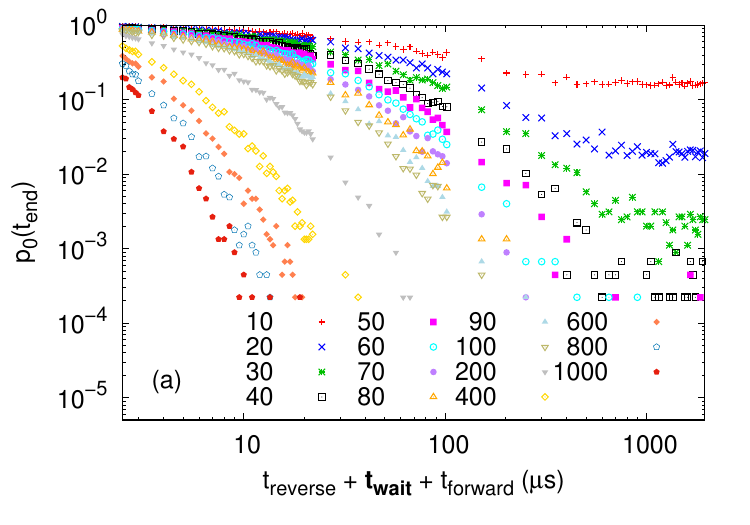}
     \end{minipage}
     \hfill
     \begin{minipage}{0.49\textwidth}
         \centering
         \includegraphics[width=\textwidth]{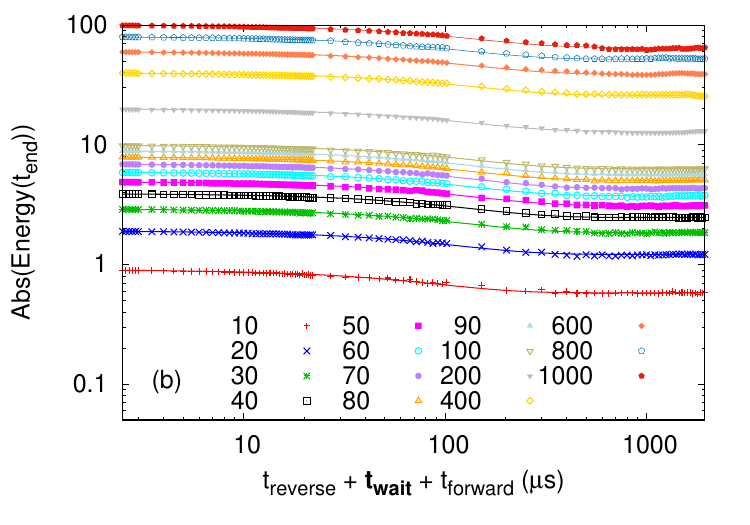}
     \end{minipage} 
     \hfill
     \caption{(Color online) Empirical data for the WTS for (a) success probability $p_0(t_{end})$ and (b) absolute values of the mean energy for ferromagnetic Ising spin chains of size $10 \leq N \leq 1000$ and $s_r=0.7$. Lines in (b) serve as the guide to the eye.}
    \label{fig:DW_size}
\end{figure*}

Figure.~\ref{fig:DW_size}(a) shows empirical data of the ground state probabilities obtained from the WTS for various problem sizes up to $N=1000$, averaged over 4500 samples.
With increasing problem size, the $p_0(2000\,\mu s)$'s saturate at decreasingly lower values, up to $N \leq 30$. For larger $N$, the required number of samples is prohibitive to make  definite statements.

Figure.~\ref{fig:DW_size}(b) shows empirical data of the absolute value of the mean energies obtained from the WTS for various problem sizes up to $N=1000$, averaged over 4500 samples.
As discussed in section~\ref{sec:DW}, the values at $t_{end}=2000\,\mu s$ are in excellent agreement with the corresponding equilibrium values of the one-dimensional ferromagnetic Ising chain.

\bibliography{references}
\end{document}